\documentclass[superscriptaddress, reprint,nofootinbib]{revtex4-2}
\usepackage{graphicx}
\graphicspath{ {../fig/}}

\usepackage{amssymb}
\usepackage{amsmath}
\usepackage{amsthm}
\usepackage{amsfonts}
\usepackage{xfrac}
\usepackage{wasysym}
\usepackage{textcomp}
\usepackage{bbm}

\usepackage{braket}
\usepackage[version = 4]{mhchem}
\usepackage{epstopdf} 
\usepackage{physics}

\usepackage{hyperref}
\usepackage{array}
\usepackage{comment}

\newcommand{\figSize}{0.49}

\begin{document}

\author{Nathan P. Giha}
\email{giha@umich.edu}
\author{Stefano Marin}
\author{James A. Baker}
\author{Isabel E. Hernandez}
\thanks{Present address: Department of Nuclear Engineering, University of California, Berkeley, CA 94720, USA}

\affiliation{Department of Nuclear Engineering and Radiological Sciences, University of Michigan, Ann Arbor, MI 48109, USA}

\author{Keegan J. Kelly}
\author{Matthew Devlin}
\author{John M. O'Donnell}

\affiliation{Physics Division, Los Alamos National Laboratory, Los Alamos, NM 87545, USA}

\author{Ramona Vogt}
\affiliation{Nuclear and Chemical Sciences Division,
Lawrence Livermore National Laboratory, Livermore, CA 94550, USA}
\affiliation{Physics and Astronomy Department,
University of California, Davis, CA 95616, USA}

\author{J\o rgen Randrup}
\affiliation{Nuclear Science Division, Lawrence Berkeley
National Laboratory, Berkeley, CA 94720, USA}

\author{Patrick Talou}
\affiliation{Computational Physics Division, Los Alamos National Laboratory, Los Alamos, NM 87545, USA}

\author{Ionel Stetcu}
\author{Amy E. Lovell}
\affiliation{Theoretical Physics Division, Los Alamos National Laboratory, Los Alamos, NM 87545, USA}

\author{Olivier Litaize}
\author{Olivier Serot}
\author{Abdelhazize Chebboubi}
\affiliation{CEA, DES, IRESNE, DER, SPRC, Physics Studies Laboratory, Cadarache, F-13108 Saint-Paul-lès-Durance, France}

\author{Ching-Yen Wu}
\affiliation{Nuclear and Chemical Sciences Division,
Lawrence Livermore National Laboratory, Livermore, CA 94550, USA}

\author{Shaun D. Clarke}
\affiliation{Department of Nuclear Engineering and Radiological Sciences, University of Michigan, Ann Arbor, MI 48109, USA}

\author{Sara A. Pozzi}
\affiliation{Department of Nuclear Engineering and Radiological Sciences, University of Michigan, Ann Arbor, MI 48109, USA}
\affiliation{Department of Physics, University of Michigan, Ann Arbor, MI 48109, USA}

\title{Correlations between energy and $\ensuremath{\mathbf{\gamma}}$-ray emission in $\ensuremath{\mathbf{^{239}Pu \vb*{(n,\mathrm{f})}}}$}

\date{\today}

\begin{abstract}
We study $\gamma$-ray emission following $^{239}\mathrm{Pu}(n,\mathrm{f})$ over an incident neutron energy range of $2 < E_i < 40$ MeV. We present the first experimental evidence for positive correlations between the total angular momentum generated in fission and the excitation energy of the compound nucleus prior to fission. The $\gamma$-ray multiplicity increases linearly with incident energy below the 2\textsuperscript{nd}-chance fission threshold with a slope of $0.085 \pm 0.010$~MeV$^{-1}$. This linear trend appears to hold for the average excitation energy of the compound nucleus between $9 < \langle E_x \rangle < 19$ MeV. Most of the multiplicity increase comes from an enhancement around a $\gamma$-ray energy of 0.7 MeV, which we interpret as stretched quadrupole $\gamma$ rays that indicate an increase in total fission-fragment angular momentum with excitation energy.

\end{abstract}


\maketitle


\section{Introduction}

Nuclear fission was discovered over eighty years ago~\cite{Hahn1939, MEITNER1939} but the microscopic details of the process are still not fully understood.  The importance of fission in the \textit{r}-process of nucleosynthesis~\cite{Goriely2015, Vassh2019, Mumpower2020,  Vassh2020, Wang2020}, synthesis of superheavy nuclei~\cite{Zagrebaev2001, Itkis2015}, and developing Generation-IV fast-fission reactors~\cite{Rimpault2012} has motivated renewed interest in predictive fission models like \textsc{cgmf}~\cite{Talou2021}, \textsc{fifrelin}~\cite{Litaize2015}, and \textsc{freya}~\cite{Vogt2009}. One of the most prominent questions in contemporary fission physics is the nature of the mechanism by which two fragments, each with 6-8 $\hbar$ of angular momentum, emerge from a system with zero or near-zero angular momentum. Recently, there has been much discussion regarding angular momentum generation in fission~\cite{Vogt2021, Wilson2021, Bulgac_FFIntrinSpins2021, RandrupVogt_GenOfFragAngMom2021, Marevic2021, Stetcu2021}. This discussion highlights the lack of definitive experimental evidence for any particular angular momentum generation mechanism. Experimentally-determined correlations between fission observables offer powerful tests of fission models and will be instrumental in discovering which mechanism is correct.

Because the nascent fission fragments quickly de-excite, it is not possible to directly measure the intrinsic angular momenta of the fragments immediately after scission~\cite{Gonnenwein2014}. This information is encoded in the subsequent fragment de-excitation via neutron and $\gamma$-ray emission. Electric quadrupole $(E2)$ transitions along yrast bands, in particular, remove most of the intrinsic angular momentum~\cite{Wilson2021, Wilhelmy1972}. Therefore, simultaneous measurements of these $E2$ $\gamma$ rays and system energy are experimentally-accessible signatures of correlations between the angular momentum and excitation energy of fission fragments. Understanding the relationship between the excitation energy of the fissioning system\textemdash and consequently of the fragments\textemdash and the fragment angular momenta is critical for constraining the possible mechanisms of angular momentum generation.
For example, the popular statistical model posits that the high angular momenta with which fragments emerge are solely due to the higher density of high-angular momentum states at large excitation energy~\cite{Moretto1989}. This model would result in a nonlinear dependence of angular momentum on excitation energy.

Experimental investigations on the dependence of $\gamma$-ray emission on the energy of the fissioning system are sparse~\cite{Frehaut1983,Frehaut1989,Qi2018,Laborie2018,Oberstedt2020,Rose2017,Gjestvang2021}. In most cases, the experiments investigated only a few different energies or a limited energy range, and could not resolve any trends as a result. Table~\ref{tab:exps} summarizes these experiments, listing the investigated reaction, energies, and whether or not they observed changes in the $\gamma$-ray multiplicity and spectrum. The ENDF/B-VIII.0 evaluation for $^{239}\mathrm{Pu}(n,\mathrm{f})$ is also included. Note that only Gjestvang~\textit{et al.}~\cite{Gjestvang2021} identified a significant change in $\gamma$-ray multiplicity. Only Laborie~\textit{et al.}~\cite{Laborie2018} found changes in the $\gamma$-ray spectrum, but exclusively above 2 MeV in $\gamma$-ray energy, uncharacteristic of $E2$ transitions.

\begingroup
\squeezetable
\begin{table}
\begin{ruledtabular}
\caption{Fission $\gamma$-ray measurements and whether they were able to statistically resolve changes in $\gamma$-ray multiplicity, $ \Delta \overline{N}_\gamma$, or changes in the $\gamma$-ray spectrum, $\Delta \text{Spec}$. For neutron-induced reactions other than $^{239}\mathrm{Pu}(n,\mathrm{f})$, $E_x$ above the 2\textsuperscript{nd}-chance fission threshold are omitted. Experiments by Fr\'ehaut are frequently cited in discussions about the energy dependence of angular momentum in fission, but the conclusions in Refs.~\cite{Frehaut1983} and \cite{Frehaut1989} are contradictory.}
\label{tab:exps}
\begin{tabular}{lccccc}
    Reference & Reaction & $E_n$ & $E_x$ & $\Delta \overline{N}_\gamma$ & $\Delta \text{Spec}$ \\
    \hline
    This work & $^{239}\mathrm{Pu}(n,\mathrm{f})$ & 2-40 & 9-19 & \checkmark & \checkmark\\
    ENDF/B-VIII.0~\cite{ENDF_BROWN20181} & $^{239}\mathrm{Pu}(n,\mathrm{f})$ & 0-20 & 6.53-19 & \checkmark & \\
    Fr\'ehaut~\cite{Frehaut1983, Frehaut1989} &  $^{235}\mathrm{U}(n,\mathrm{f})$ & 1.14-14.66 & 7.69-12.22 & N/A & N/A\\
    Qi~\cite{Qi2018} & $^{238}\mathrm{U}(n,\mathrm{f})$ & 1.90,4.90 & 6.71,9.61 &  & \\
    Laborie~\cite{Laborie2018} & $^{238}\mathrm{U}(n,\mathrm{f})$ & 1.6,5.1,15.0 & 6.41,9.91 &  & \checkmark\\
    Oberstedt~\cite{Oberstedt2020} & $^{235}\mathrm{U}(n,\mathrm{f})$ & $\overline{E}_n = 1.7$ & $\overline{E}_x = 8.25$ &  & \\
    Rose~\cite{Rose2017} & $^{233}\mathrm{U}(d,p\mathrm{f})$ & - & 4.8-10 &  & \\
    Rose~\cite{Rose2017} & $^{239}\mathrm{Pu}(d,p\mathrm{f})$ & - & 4.5-8.8 &  & \\
    Gjestvang~\cite{Gjestvang2021} & $^{240}\mathrm{Pu}(d,p\mathrm{f})$  & - & 5.5-8.5 & \checkmark & 
\end{tabular}
\end{ruledtabular}
\end{table}
\endgroup

In this paper we analyze the $^{239}\mathrm{Pu}(n,\mathrm{f})$ data from Kelly~\textit{et al.}~\cite{Kelly_PFNS_2020}, in which a broad range of excited states of $^{240}\mathrm{Pu}^*$ were populated. We present clear experimental evidence for increasing $\gamma$-ray multiplicity, $\overline N_\gamma$, over the incident neutron energy range of $2 < E_i < 40$ MeV. We find an approximately linear relationship between $\overline N_\gamma$ and the average compound nucleus excitation energy, $\langle E_x \rangle$, within $9 < \langle E_x \rangle < 19$ MeV. Furthermore, by differentiating with respect to the $\gamma$-ray energy, $E_\gamma$, we find the $\gamma$-ray multiplicity around $E_\gamma = 0.7$ MeV\textemdash characteristic of $E2$ transitions along fragment rotational bands\textemdash increases with the excitation energy of the compound system. We ultimately suggest a positive, approximately linear angular momentum-energy correlation in the measured energy range.

\section{Experiment and Analysis}

The experiment was carried out at the Los Alamos Neutron Science Center~\cite{lansce}, where a broad-spectrum neutron beam was produced via spallation reaction of an 800 MeV proton beam on a tungsten target. The neutron beam was incident on a multi-foil Parallel-Plate Avalanche Counter (PPAC)~\cite{ppac_2015} containing $100$ mg of $^{239}\mathrm{Pu}$, 21.5 m from the spallation target. Neutron-induced fission was measured in the PPAC and the neutrons and $\gamma$ rays emitted by the fragments were measured using the Chi-Nu liquid scintillator array, a hemispherical array of 54 EJ-309~\cite{ej309} organic scintillator detectors. We separate the data into quasi-monoenergetic bins of incident energy, $E_i$, determined by the neutron time of flight between spallation and measurement of fission in the PPAC. A detailed description of the experiment that generated these data is available in Ref.~\cite{Kelly_PFNS_2020}. Whereas Kelly~\textit{et al.} focused on prompt fission neutron measurements, we apply an entirely new analysis to the $\gamma$-ray data.

\subsection{Analysis}
Fission $\gamma$ rays and neutrons, measured in coincidence with beam and PPAC triggers, are discriminated based on pulse shape and time of flight. After applying both discrimination techniques, particle misclassification becomes negligible~\cite{Marin2020}. We collect $\gamma$ rays within a window of 5 ns before to 10 ns after the PPAC trigger. The full width at half maximum of this coincidence peak is 3.1 ns. To recover the emitted $\gamma$-ray features from the detected events, several corrections are applied. Since the target nucleus $^{239}\mathrm{Pu}$ is unstable to $\alpha$ decay, the PPAC signal from pileup of multiple $\alpha$ events cannot always be separated from that produced by decelerating fission fragments. The bias associated with erroneous triggers from $^{239}\mathrm{Pu}$ $\alpha$ decay is estimated by examining the measured PPAC activity and spectrum in the absence of beam.

We quantify the effect of chance coincidences between the $\gamma$-ray background and the beam trigger by introducing a random coincidence signal in the analysis. Its contribution is small and we subtract it. While multiple $\gamma$ rays and neutrons are usually emitted in the same fission, pileup can be neglected due to the low absolute efficiency of the detector array: about $2.9\%$.

The pulsed nature of the broad-spectrum neutron beam results in low-energy neutrons from a beam micropulse arriving at the target simultaneously with high-energy neutrons from the next micropulse. We estimate the amount of fission induced by these low-energy neutrons and subtract. This correction is negligible at low $E_i$ and never exceeds $3.4\%$ as $E_i$ approaches 40 MeV.

\begin{figure}[htb!]
    \centering
    \includegraphics[width = \figSize\textwidth]{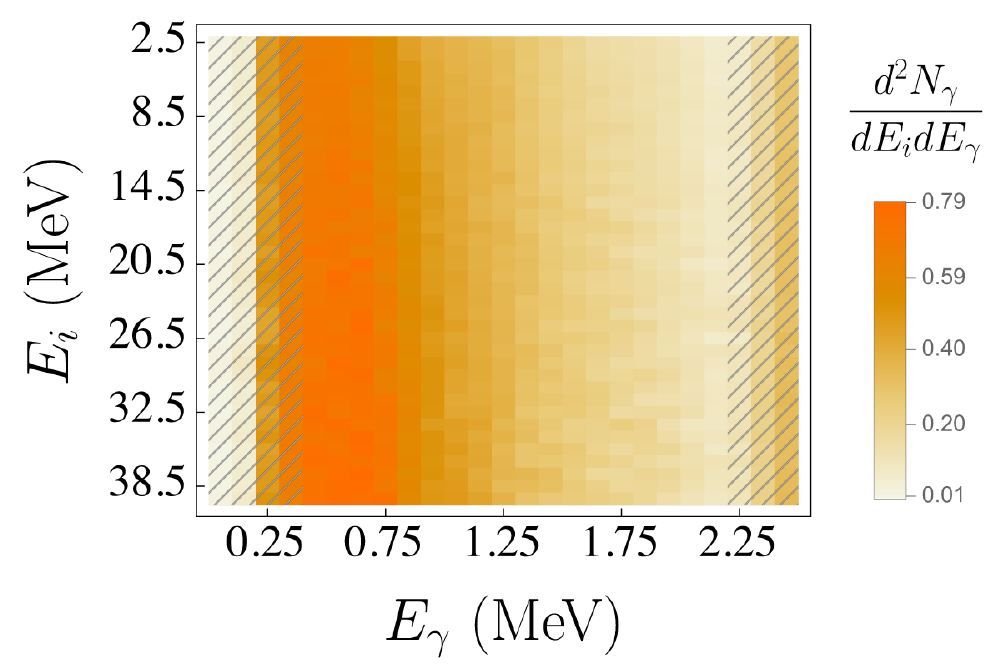}
    \caption{Measured $\gamma$-ray spectra for each quasi-monoenergetic incident neutron energy bin, $E_i$. The hatched regions fall outside of the $E_\gamma$ acceptance window.}
    \label{fig:specs}
\end{figure}

Finally, we apply the following unfolding procedure to recover the emitted $\gamma$-ray spectrum at each $E_i$: we first model the system response of the Chi-Nu liquid scintillator array using isotropic, monoenergetic photon sources in \textsc{mcnpx-polimi}~\cite{Pozzi2003}. We then convolve the resulting response matrix with experimentally-determined detector resolution and a scintillator light output threshold of 0.1 MeVee, and then invert it via Tikhonov regularization~\cite{Schmitt2016}. This procedure corrects the measured multiplicity for efficiency and unfolds the emitted $E_\gamma$ spectrum from the measured $\gamma$-ray light output spectrum. The measured $\gamma$-ray spectra for each $E_i$ bin are shown in Fig.~\ref{fig:specs}. The energy resolution, including both detector resolution and uncertainty introduced by the unfolding procedure, is $\approx 19\%$ in the analyzed $\gamma$-ray energy range. By comparing the unfolded $\gamma$-ray spectrum at our lowest energy bin, $2 < E_i < 3$ MeV, with the ENDF/B-VIII.0 evaluated spectrum for $^{239}\mathrm{Pu}(n_{\mathrm{th}}, \mathrm{f})$~\cite{Stetcu2020}, we determined that the unfolding procedure reproduced the correct spectral shape and magnitude between $0.4 < E_\gamma < 2.2$ MeV. This limitation is reflected in Fig.~\ref{fig:specs}, where the hatched regions fall outside the acceptance window.

The $\overline N_\gamma$ reported throughout this paper thus includes only gamma rays within this acceptance window of $0.4 < E_\gamma < 2.2$ MeV, representing $\approx60\%$ of the integrated $^{239}\mathrm{Pu}(n_{\mathrm{th}}, \mathrm{f})$ $\gamma$-ray spectrum above 0.1 MeV. Almost all of the excluded $\gamma$ rays fall below the acceptance region. We estimate the unfolding uncertainty in $\overline N_\gamma$ by constructing a covariance matrix by varying the regularization parameter.

\subsection{Fission codes}
The fission models \textsc{cgmf}~\cite{Talou2021}, \textsc{fifrelin}~\cite{Litaize2015}, and \textsc{freya}~\cite{Vogt2009} were employed to examine how different treatments of fragment formation and particle emission affect the relationship between $\gamma$-ray emission and incident energy. All three codes in this manuscript use phenomenological models and while the underlying principles are sometimes similar, varying treatments of determining the initial fragment properties and their subsequent de-excitation can result in very different predictions of the $\gamma$-ray spectrum and multiplicity. We provide short descriptions of each model here, and point the reader to suitable references for more details.

\subsubsection{CGMF~\cite{Talou2021}}
\textsc{cgmf} takes as input the pre-neutron fission fragment mass and kinetic energy distributions and samples from these distributions to determine the total excitation energy of the fragments. This total excitation energy is shared between the fragments based on a mass-dependent nuclear temperature ratio law. The angular momentum of each fragment is subsequently sampled from a spin distribution closely following Bethe's work~\cite{Bethe1936}, with a spin cut-off parameter (called $B^2$ in Ref.~\cite{Talou2021}) that depends on the moment of inertia of the fragment and is proportional to the fragment temperature. Note that $B^2$ includes an adjustable scaling factor that depends linearly on $E_i$ and is used to tune the competition between neutrons and photons to fit experimental photon data. $\textsc{cgmf}$ handles pre-fission neutron emission using probabilities calculated with the \textsc{CoH}$_3$ code~\cite{Kawano2019}.

\textsc{cgmf} implements the Hauser-Feshbach statistical nuclear reaction model to follow the de-excitation of fission fragments. It uses a spherical optical model potential to determine neutron transmission coefficients. $\gamma$-ray transmission coefficients are determined using the strength function formalism, where the continuum level density follows the Fermi-gas formula at high excitation energies and a constant-temperature formula at lower excitation energies. Discrete levels are imported from the RIPL-3~\cite{Capote2009} database where available. More details on the specific models used, as well as a complete list of the input files required to run \textsc{cgmf}, are available in Table 2 of Ref.~\cite{Talou2021}.

\subsubsection{FIFRELIN~\cite{Litaize2015}}
Similarly to \textsc{cgmf}, the pre-neutron fission fragment mass and kinetic energy distributions are used as inputs in \textsc{fifrelin} and sampled, in order to calculate the total excitation energy of the fragments. \textsc{fifrelin} also employs an empirical mass-dependent temperature ratio of the fragments to partition the excitation energy between them, and the total angular momentum of each fragment is statistically sampled following Bethe's work. Different models for the spin cut-off parameter can be used~\cite{Thulliez2016}; in the Inertia+Shell model used in this work, the spin cut-off depends on the mass, ground-state deformation, and temperature of the nucleus as well as shell effects. This model includes one free scaling parameter that is allowed to vary with $E_x$. 
Note that in \textsc{fifrelin}, the four free parameters are adjusted to reproduce the total prompt neutron multiplicity in the JEFF-3.3 library~\cite{Plompen2020}. In other words, there is no explicit dependence on experimental $\gamma$-ray data, including in the spin cut-off scaling parameter. \textsc{fifrelin} does not include pre-fission neutron emission.

\textsc{fifrelin} implements a coupled Hauser-Feshbach algorithm based on the concept of Nuclear Realization, established by Becvar~\cite{Becvar1998} and implemented by Regnier~\textit{et al.}~\cite{Regnier2016} for neutron/$\gamma$/electron coupled emission from an excited nucleus. Neutron transmission coefficients are governed by optical model calculations. $\gamma$-ray emission is determined by the strength function formalism. Somewhat uniquely, in each realization an artificial set of levels is generated based on expected level densities, and the partial widths of a given transition energy are allowed to fluctuate~\cite{Regnier2016, Litaize2017}. This strategy is potentially important for modeling $\gamma$-ray observables when the input nuclear structure data files are deficient~\cite{Litaize2018InfluenceStructure}.

\subsubsection{FREYA~\cite{Vogt2009}}

Just as in the previously mentioned codes the mass, charge, and total kinetic energy distributions of the fragments are sampled at the beginning of a fission event in \textsc{freya}. The temperature sharing is directly specified by a free parameter. The angular momenta of the fragments in \textsc{freya} are generated based on the ``spin temperature,'' $T_S$, which is the temperature of the dinuclear system at scission multiplied by a free parameter, $c_S$. In \textsc{freya}, this free parameter does not depend on energy. Contributions from the dinuclear rotational modes available at scission\textemdash tilting, twisting, wriggling, and bending\textemdash are statistically populated based on this spin temperature~\cite{Vogt2021}. This is in contrast to the previous two models, which sample the fragment angular momenta based on the nascent fragment temperatures after they are separated. Prefission neutron emission is treated the same way as postfission neutron evaporation from the fragments.

The fragments de-excite via neutron evaporation with a black-body spectrum until the available intrinsic energy falls below the neutron separation energy. Statistical photons are then emitted with a black-body spectrum modulated by a giant dipole resonance form factor. In \textsc{freya}, all statistical photons remove 1 $\hbar$ of angular momentum. Once the excitation energy is sufficiently low, evaluated discrete transitions from the  RIPL-3 data library~\cite{Capote2009} are used until the ground state or a sufficiently long-lived isomeric state is reached~\cite{Randrup2014}. The free parameters in \textsc{freya} are summarized in Ref.~\cite{Vogt2017}.

\section{Results}

\begin{figure}[htb!]
    \centering
    \includegraphics[width = \figSize\textwidth]{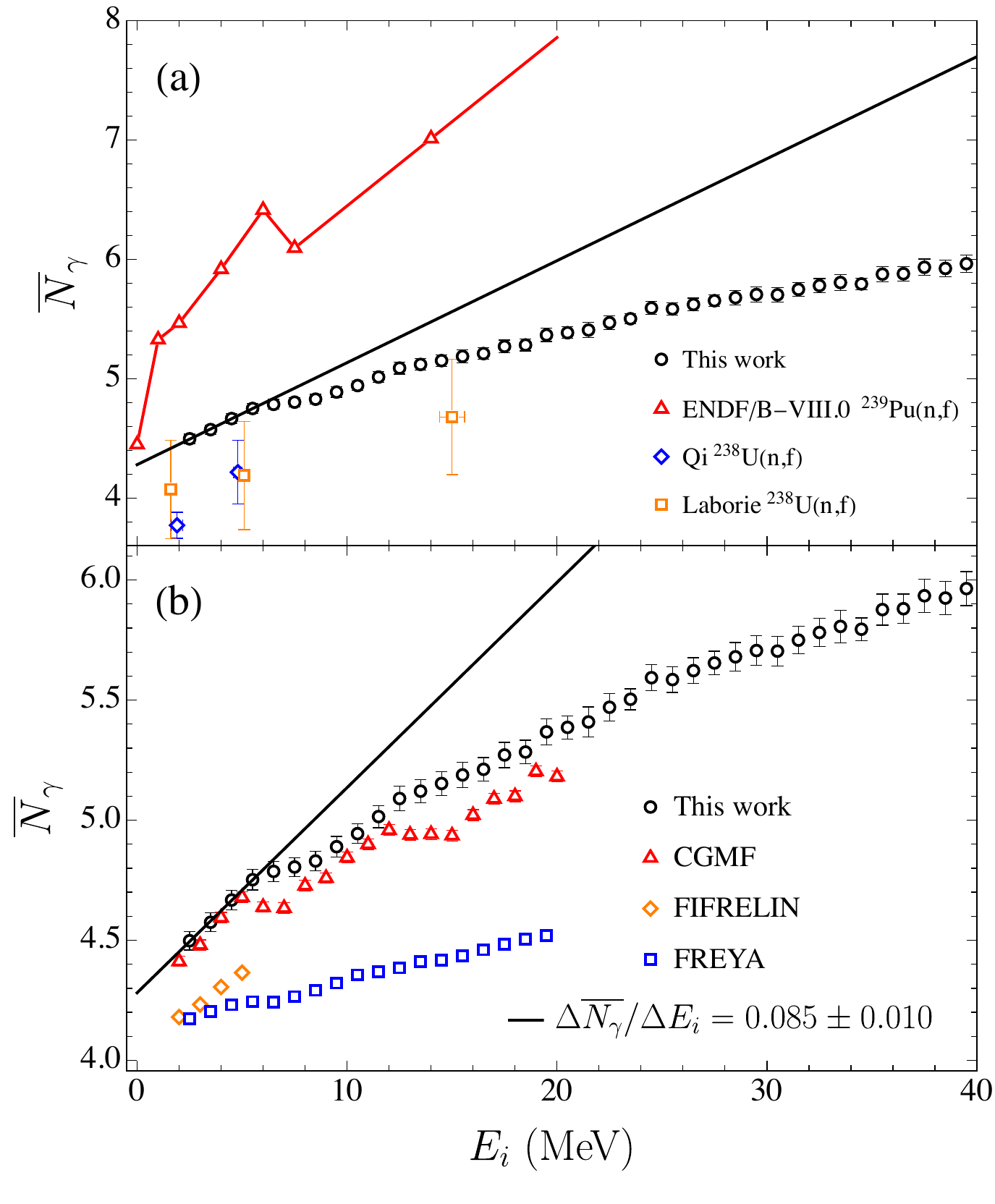}
    \caption{$\overline N_\gamma$ between $0.4 < E_\gamma < 2.2$ MeV as a function of $E_i$ for $2 < E_i < 40$ MeV. Data where $E_i$ is below the $^{240}\mathrm{Pu}$ inner fission barrier height, $B_\mathrm{f} = 6.05$ MeV~\cite{Capote2009}, are fit with a black line. The bin width is 1 MeV.}
    \label{fig:multVsEi}
\end{figure}

In Fig.~\ref{fig:multVsEi}, we present the relationship between $\overline N_\gamma$ and $E_i$ between $2 < E_i < 40$ MeV. Our data show a clear increase in $\overline N_\gamma$ across the entire $E_i$ range. Uncertainties include variation across PPAC foils and unfolding; statistical uncertainties are comparatively negligible. Also plotted in Fig.~\ref{fig:multVsEi}(a) are $\gamma$-ray multiplicities from the ENDF/B-VIII.0 evaluation~\cite{ENDF_BROWN20181} and data from Qi~\cite{Qi2018} and Laborie~\cite{Laborie2018}. These data are scaled down to match our $0.4 < E_\gamma < 2.2$ MeV acceptance region. We integrate the ENDF/B-VIII.0 $^{239}\mathrm{Pu}(n,\mathrm{f})$ and $^{238}\mathrm{U}(n,\mathrm{f})$ $\gamma$-ray spectra within our acceptance range, then again for a threshold $E_\gamma > 0.1$ MeV. Most of the experimental results are reported for a 0.1 MeV threshold and extend up to sufficiently high $E_\gamma$ that their upper limit does not significantly affect $\overline N_\gamma$. Thus, the evaluation and experimental data in Fig.~\ref{fig:multVsEi} are scaled down by the ratio of these two integrals for the appropriate reaction. Even with this correction, we do not necessarily expect the Qi~\cite{Qi2018} and Laborie~\cite{Laborie2018} data to agree with our data since they study a different reaction. The ENDF/B-VIII.0 points above thermal fission were inferred from total $\gamma$-ray production data, assuming a 20\% uncertainty~\cite{Stetcu2020}.


We note that $\overline N_\gamma$ varies linearly with $E_i$ below the $2^{\mathrm{nd}}$-chance fission threshold with a slope of $\Delta \overline{N}_\gamma / \Delta E_i = 0.085\pm0.010$~MeV$^{-1}$. This behavior was also observed by Gjestvang~\textit{et al.} in $^{240}\mathrm{Pu}(d,p\mathrm{f})$, where they found a slope of $0.08\pm0.03$~MeV$^{-1}$. Extrapolating this fit down to $E_i = 0$ yields good agreement with the well-studied multiplicity at thermal fission~\cite{Gatera2017}. Uncertainty on the slope includes variation across PPAC foils, uncertainty from unfolding, and estimated variance of the fitted slope. 

In Fig.~\ref{fig:multVsEi}(b), we compare our data to predictions from \textsc{fifrelin} and the release versions of \textsc{cgmf} and \textsc{freya} for $\overline N_\gamma$ within the acceptance window as a function of $E_i$. Only data below the second-chance fission threshold are shown for \textsc{fifrelin}, since it does not include pre-fission emission. \textsc{cgmf} predicts a similar trend, although the discontinuities at the $n$\textsuperscript{th}-chance fission thresholds are overemphasized compared to experiment. \textsc{freya} predicts about 0.5 too few $\gamma$ rays within the acceptance region. The model uncertainties are statistical.

The neutron separation energies, $S_n$, of different fissioning isotopes can vary by several MeV so comparing $\gamma$-ray emission from different reactions at a given $E_i$ is not always appropriate. It is instructive to instead look at the excitation energy of the fissioning nucleus, $E_x$, which is independent of this variation. If we neglect the small kinetic energy imparted to the compound nucleus by the incident neutron, the excitation energy of the pre-fission $^{240}\mathrm{Pu}^*$ nucleus is 
\begin{equation}
    E_x = E_i + S_n^{(240)},
    \label{eq:Ex1}
\end{equation}
where $E_i$ is the incident neutron energy and $S_n^{(240)} = 6.53$ MeV is the neutron separation energy of the compound $^{240}\mathrm{Pu}^*$ nucleus. However, the $E_x$\textemdash and in fact, the isotope\textemdash of the compound nucleus just before fission cannot be uniquely determined once the incident neutron energy exceeds the fission barrier height, $B_\mathrm{f}$, due to the presence of multi-chance fission and pre-equilibrium neutron emission. Thus, multiple $E_x$ values are possible for a given $E_i > B_\mathrm{f}$ and the average excitation energy, $\langle E_x \rangle$, of the fissioning nucleus is generally lower than what may be expected from Eq.~\eqref{eq:Ex1}. At a fixed $E_i$, $\langle E_x \rangle$ can be written

\begin{equation}
\langle E_x \rangle = E_i + S_n^{(240)} - \sum_{j = 1}  \left[ S_n^{(240-j+1)} + \langle k_j \rangle \right]p_j
\label{eq:Ex2}
\end{equation}
where $S_n^{(240-j+1)}$ is the separation energy of the $j$\textsuperscript{th} neutron, $\langle k_j \rangle \equiv \langle k_j \rangle (E_i)$ is the average kinetic energy of the $j$\textsuperscript{th} pre-fission neutron, and $p_j \equiv p_j (E_i)$ is the probability of emitting $j$ neutrons prior to fission. Note that Pu isotopes lighter than $^{240}\mathrm{Pu}^*$ contribute to the total observed fissions when prefission neutron emission occurs. For compound nuclei that are close in mass, correlations between $\langle E_x \rangle$ and $\gamma$ rays should be relatively independent of the isotope. $\langle k_j \rangle$ and $p_j$ are model dependent; $\langle k_j \rangle$ was estimated using \textsc{cgmf} and $p_j$ was calculated using the ENDF/B-VII.1 cross sections~\cite{Chadwick2011}. We do not consider pre-equilibrium $\gamma$-ray emission since neutron-$\gamma$ competition is minimal when $E_x$ is high enough for pre-fission processes to occur~\cite{Mumpower2016, Spyrou2016}.

\begin{figure}[htb!]
    \centering
    \includegraphics[width = \figSize\textwidth]{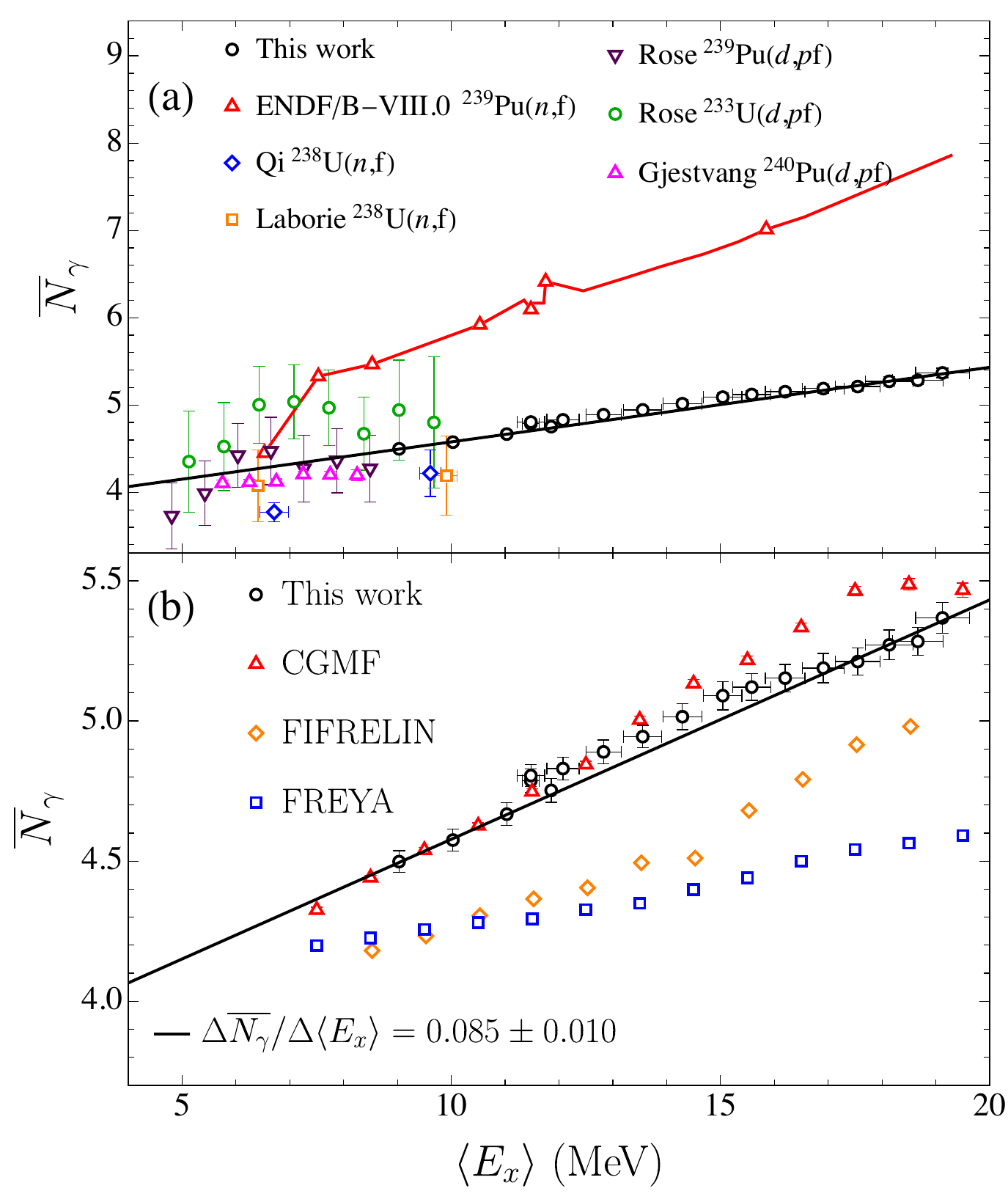}
    \caption{$\overline N_\gamma$ between $0.4 < E_\gamma < 2.2$ MeV as a function of $\langle E_x \rangle$ for $9 < \langle E_x \rangle < 19$ MeV. The black line is the same as in Fig.~\ref{fig:multVsEi}, shifted to the right by $S_n^{(240)}$, see Eq.~\eqref{eq:Ex1}.
    }
    \label{fig:multVsEx}
\end{figure}

$E_x$ becomes a better description for the state of the compound nucleus just before fission once $E_i > B_\mathrm{f}$. To investigate the relationship between $\overline N_\gamma$ and $E_x$, in Fig.~\ref{fig:multVsEx} we translate $E_i$ to $\langle E_x \rangle$ using Eq.~\eqref{eq:Ex2}. This translation corrects for the effects introduced by pre-fission neutron emission and reveals the approximate linearity of $\overline N_\gamma$ with respect to $\langle E_x \rangle$ for  $9 < \langle E_x \rangle < 19$ MeV. The model-dependent parameters $p_j$ and $\langle k_j \rangle$ in Eq.~\eqref{eq:Ex2} bias the translation, so we assign 10\% uncertainties to $p_j$ and $\langle k_j \rangle$ which give rise to the horizontal uncertainties on our data. The models do not predict these values for $E_i > 20$ MeV, so the data above this limit are excluded from Fig.~\ref{fig:multVsEx}.

Also plotted in Fig.~\ref{fig:multVsEx}(a) are the ENDF/B-VIII.0 evaluation~\cite{ENDF_BROWN20181} and the Qi~\cite{Qi2018}, Laborie~\cite{Laborie2018}, Rose~\cite{Rose2017}, and Gjestvang~\cite{Gjestvang2021} data. The energy transformation in Eq.~\eqref{eq:Ex2} was also applied to the ENDF/B-VIII.0 evaluation. The incident energies of Qi and Laborie are shifted using Eq.~\eqref{eq:Ex1} with the appropriate $S_n$ for each reaction. The $E_i=15.0$ MeV point from Laborie is omitted due to lack of nuclear data for determining $p_j$ and $\langle k_j \rangle$ for $^{238}\mathrm{U}(n,\mathrm{f})$. 

Our data agree well with other experiments in the limited range of overlap, although agreement with our extrapolation to lower $E_x$ is mixed. We note in the cases of Rose~\cite{Rose2017} and Gjestvang~\cite{Gjestvang2021} that some disagreement could arise from ion-induced fission populating different states of the compound nucleus~\cite{Boutoux2012,Zeiser2019}. Recent theoretical work~\cite{Vogt2021}, however, concluded that the angular momentum of the compound nucleus has little effect on the angular momenta of the fragments, which would decouple the $\gamma$-ray multiplicity from the choice of reaction used to form the compound nucleus.


In Fig.~\ref{fig:multVsEx}(b) we compare our data to predictions from \textsc{cgmf}, \textsc{fifrelin}, and \textsc{freya} for $\overline{N}_\gamma$ within $0.4 < E_\gamma < 2.2$ MeV as a function of $E_x$. In \textsc{cgmf} and \textsc{freya}, simulated neutron-induced fission events were binned by compound nucleus excitation energy. The excitation energy of the compound nucleus was directly specified in \textsc{fifrelin}. Since \textsc{fifrelin} does not include pre-fission neutron emission, multi-chance fission does not occur and only $^{240}\mathrm{Pu}^*$ nuclei contribute. \textsc{cgmf} predicts the $\overline N_\gamma$ well across the entire $\langle E_x \rangle$ range\textemdash with some deviation at high $\langle E_x \rangle$, where we expect the energy translation in Eq.~\eqref{eq:Ex2} be more uncertain.

\textsc{cgmf} agrees quite well across most of the energy range. \textsc{fifrelin} predicts the trend well, although the absolute multiplicity within the acceptance region is too low by about 0.5 $\gamma$~rays. \textsc{freya} underestimates the positive trend and multiplicity within our acceptance window, although it still predicts positive correlations. Statistical model uncertainties are shown, although they are smaller than the markers.

\begin{figure}[htb!]
    \centering
    \includegraphics[width = \figSize\textwidth]{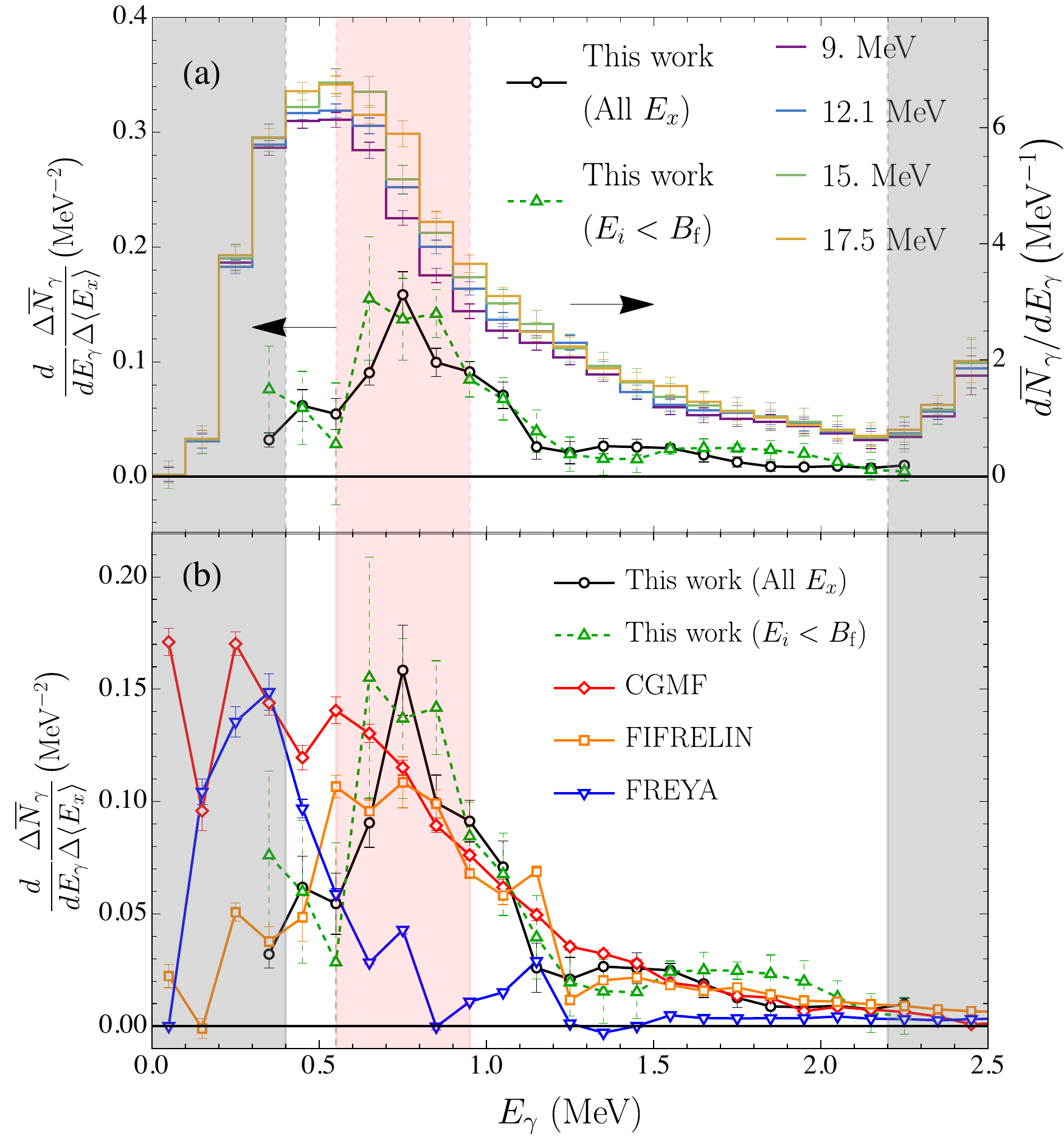}
    \caption{Dependence of the slope, $\Delta \overline{N}_\gamma / \Delta \langle E_x \rangle$, on $E_\gamma$. In (a), $\gamma$-ray spectra from the experiment for $\langle E_x \rangle = 9$, $12.1$, $15$, and $17.5$ MeV are also shown on the right-hand side. The area outside the $E_\gamma$ acceptance region is shown as the grey shaded region. $E_\gamma$ bins are 0.1 MeV.}
    \label{fig:slope}
\end{figure}

We further characterize the additional $\gamma$ rays we observe by examining how the spectrum changes with increasing $\langle E_x \rangle$. We fix $E_\gamma$ and determine the slope of a linear fit to $\overline N_\gamma$ with respect to $\langle E_x \rangle$, or $\Delta \overline N_\gamma / \Delta \langle E_x \rangle$, plotted in Fig.~\ref{fig:slope}(a). The slopes of fits to the entire $\langle E_x \rangle$ range are plotted for each $E_\gamma$, as well as fits to just the data below the 2\textsuperscript{nd}-chance fission threshold, $E_i < B_\mathrm{f}$, to provide a model-independent comparison. The uncertainties include unfolding uncertainty propagated from the covariance matrix and standard fit-parameter uncertainties. We note a particular enhancement around $E_\gamma=0.7$ MeV, characteristic of $E2$ yrast transitions in the mass range of both light and heavy fragments. This enhancement accounts for the majority of the overall increase in $\overline N_\gamma$ with respect to $\langle E_x \rangle$, suggesting most of the additional $\gamma$ rays observed at higher energies in Figs.~\ref{fig:multVsEi} and~\ref{fig:multVsEx} are $E2$ yrast transitions and remove $2\hbar$ of angular momentum each. The measured $\gamma$-ray spectra for a few $\langle E_x \rangle$ values are also plotted in Fig.~\ref{fig:slope}(a) using the right axis.

In Fig.~\ref{fig:slope}(b), slopes from fits to models are shown for comparison. The model uncertainties are standard fit-parameter uncertainties. \textsc{cgmf} agrees somewhat around the enhancement, but does not predict the dip around $E_\gamma = 0.5$ MeV that we observe in our data. We observe good agreement with \textsc{fifrelin} using the Inertia+Shell spin cut-off model, which correctly predicts the magnitude of the enhancement around $E_\gamma = 0.7$ MeV. 
\textsc{freya} does not predict the observed enhancement around $E_\gamma = 0.7$ MeV. Most of the additional $\gamma$ rays that it predicts lie below our acceptance region, explaining the discrepancy between \textsc{freya} and our data in Figs.~\ref{fig:multVsEi}(b) and \ref{fig:multVsEx}(b). We believe that \textsc{fifrelin} agrees well partially because of its nuclear realization methodology, as it creates artificial levels in nuclei where compiled discrete level libraries like RIPL~\cite{Capote2009} are lacking.


\section{Discussion}
 To draw physical conclusions, we discuss the differences between models that cause \textsc{fifrelin} to agree well with our experimental data in Fig.~\ref{fig:slope}. It is clear from this agreement that the energy-dependent spin distribution is one component of an accurate prediction. In contrast, resampling stages in \textsc{freya} eliminate the correlations between fragment excitation energy and the dinuclear temperature that is used to calculate the fragment angular momenta. The disagreement between this experiment and \textsc{freya} could be due to this decoupling of angular momentum and energy, although other differences in the models could contribute. \textsc{cgmf}'s method for calculating the spin cut-off parameter is similar to that of \textsc{fifrelin}; the spin cut-off depends on the fragment's temperature and ground-state moment of inertia in the same way in both codes. The two agree well in magnitude around the enhancement, with the main difference being that \textsc{cgmf} predicts more low-energy $\gamma$ rays while \textsc{fifrelin} and our experiment decrease at lower $E_\gamma$. Differences could arise from how the free scaling parameter is chosen. Free parameters in \textsc{fifrelin} are chosen solely to match experimental total neutron multiplicity data, while the spin cut-off scaling parameter in \textsc{cgmf} is fitted to total $\gamma$-ray energy and multiplicity data~\cite{Talou2021}. Given the similarity of their treatment, \textsc{fifrelin}'s implementation of the Nuclear Realizations established by Becvar~\cite{Becvar1998} could lead to more realistic modeling of discrete transitions in fragments with uncertain level schemes, and thus explain the better agreement at low $E_\gamma$. \textsc{freya}'s methodology for selecting the initial spin of fragments is fundamentally different, although it results in similar average spin values. Recent work regarding the angular distribution of statistical $\gamma$ rays~\cite{Marin2022} suggests that these transitions are not always stretched, and thus \textsc{freya}'s treatment may lead to a reduction in fragment spin post-statistical emission. This effect could lead to the observed deficiency in yrast $\gamma$ rays.

The fragment yield distribution also changes with excitation energy, and must be discussed.  We examined the distribution of yrast $\gamma$-ray energies as a function of the changing fragment yield to determine whether the energy threshold could bias our results. We used $E_x$-dependent fragment yields from \textsc{fifrelin} and discrete level libraries from NuDat 3.0~\cite{NuDat} to produce yield-weighted $E_\gamma$ spectra for yrast band transitions. We found that the average energy of yrast transitions with certain initial spin values, such as $8^+\rightarrow 6^+$ transitions, increases as $E_x$ increases and fragment mass yield becomes more symmetric. However, these $8^+\rightarrow 6^+$ transitions still lie within the $E_\gamma$ acceptance region at low $E_x$, so we do not suspect the $\overline{N}_\gamma$ increase around $E_\gamma = 0.7$ MeV is due to the changing fragment yields. This conclusion is consistent with our agreement with \textsc{fifrelin} (Inertia+Shell), where we can examine specific fragments and observe positive correlations between the number of yrast band transitions, and $E_x$.

\section{Conclusion}
We have presented the first direct measurement of $\gamma$-ray multiplicity, $\overline N_\gamma$, for fast neutron-induced fission of $^{239}\mathrm{Pu}$, across a large incident neutron energy range, $2 < E_i < 40 $ MeV. We observe a clear increase in $\overline N_\gamma$ over the entire range. We find an approximately linear relationship between $\overline N_\gamma$ and $E_i$ below the 2\textsuperscript{nd}-chance fission threshold, with a slope of $0.085 \pm 0.010$~MeV$^{-1}$. This relationship is preserved upon translating incident neutron energy to compound nucleus excitation energy in the range $9 < \langle E_x \rangle < 19$~MeV. These extra $\gamma$ rays are found around energies characteristic of stretched electric quadrupole transitions, experimentally confirming positive correlations between the excitation energy of the compound nucleus and the total angular momenta of the fragments. This assertion is supported by comparisons with fission model calculations. While the trend appears linear in this $E_x$ range, it is not necessarily incompatible with the statistical model of angular momentum generation. A larger range in $E_x$, particularly lower in energy, must be explored to determine the functional form. 

In future experiments, we plan to probe lower $E_x$, which will be more sensitive to the functional form of the angular momentum dependence, by examining the relationship between $\gamma$-ray emission from $^{252}\mathrm{Cf(sf)}$ and fragment mass, as well as total kinetic energy. We also suggest induced-fission experiments with higher-resolution $\gamma$-ray detectors to resolve the low-energy region of the $E_\gamma$ spectrum, as well as unambiguously identify known $E2$ transitions on an event-by-event basis. Such experiments will provide comparatively model-independent correlations between the spin distributions of fragments post-statistical emission, and their masses and excitation energies.

\acknowledgments
N.P.G. and S.M. thank the Chi-Nu experimental group at LANSCE for sharing the experimental data used i(n this analysis. N.P.G. is supported by the National Science Foundation Graduate Research Fellowship Program under Grant No. DGE 1256260. Any opinions, findings, and conclusions or recommendations expressed in this material are those of the author(s) and do not necessarily reflect the views of the National Science Foundation. N.P.G., S.M., J.A.B., I.E.H., S.D.C., and S.A.P. were funded in-part by the Consortium for Monitoring Technology and Verification under Department of Energy National Nuclear Security Administration award number DE-NA0003920. K.J.K, M.D., J.M.O., and C.Y.W. were supported by the U.S. Department of Energy through Los Alamos National Laboratory and Lawrence Livermore National Laboratory. Los Alamos National Laboratory is operated by Triad National Security, LLC, for the National Nuclear Security Administration of the U.S. Department of Energy (Contract No. 89233218CNA000001). The work of R.V. was performed under the auspices of the U.S. Department of Energy by Lawrence Livermore National Laboratory under Contract DE-AC52-07NA27344. J.R. acknowledges support from the Office of Nuclear Physics in the U.S. Department of Energy under Contract DE-AC02-05CH11231.

\bibliography{mybib}

\begin{thebibliography}{57}%
\makeatletter
\providecommand \@ifxundefined [1]{%
 \@ifx{#1\undefined}
}%
\providecommand \@ifnum [1]{%
 \ifnum #1\expandafter \@firstoftwo
 \else \expandafter \@secondoftwo
 \fi
}%
\providecommand \@ifx [1]{%
 \ifx #1\expandafter \@firstoftwo
 \else \expandafter \@secondoftwo
 \fi
}%
\providecommand \natexlab [1]{#1}%
\providecommand \enquote  [1]{``#1''}%
\providecommand \bibnamefont  [1]{#1}%
\providecommand \bibfnamefont [1]{#1}%
\providecommand \citenamefont [1]{#1}%
\providecommand \href@noop [0]{\@secondoftwo}%
\providecommand \href [0]{\begingroup \@sanitize@url \@href}%
\providecommand \@href[1]{\@@startlink{#1}\@@href}%
\providecommand \@@href[1]{\endgroup#1\@@endlink}%
\providecommand \@sanitize@url [0]{\catcode `\\12\catcode `\$12\catcode
  `\&12\catcode `\#12\catcode `\^12\catcode `\_12\catcode `\%12\relax}%
\providecommand \@@startlink[1]{}%
\providecommand \@@endlink[0]{}%
\providecommand \url  [0]{\begingroup\@sanitize@url \@url }%
\providecommand \@url [1]{\endgroup\@href {#1}{\urlprefix }}%
\providecommand \urlprefix  [0]{URL }%
\providecommand \Eprint [0]{\href }%
\providecommand \doibase [0]{http://dx.doi.org/}%
\providecommand \selectlanguage [0]{\@gobble}%
\providecommand \bibinfo  [0]{\@secondoftwo}%
\providecommand \bibfield  [0]{\@secondoftwo}%
\providecommand \translation [1]{[#1]}%
\providecommand \BibitemOpen [0]{}%
\providecommand \bibitemStop [0]{}%
\providecommand \bibitemNoStop [0]{.\EOS\space}%
\providecommand \EOS [0]{\spacefactor3000\relax}%
\providecommand \BibitemShut  [1]{\csname bibitem#1\endcsname}%
\let\auto@bib@innerbib\@empty
\bibitem [{\citenamefont {Hahn}\ and\ \citenamefont
  {Strassmann}(1939)}]{Hahn1939}%
  \BibitemOpen
  \bibfield  {author} {\bibinfo {author} {\bibfnamefont {O.}~\bibnamefont
  {Hahn}}\ and\ \bibinfo {author} {\bibfnamefont {F.}~\bibnamefont
  {Strassmann}},\ }\href {\doibase 10.1007/BF01488241} {\bibfield  {journal}
  {\bibinfo  {journal} {Naturwissenschaften}\ }\textbf {\bibinfo {volume}
  {27}},\ \bibinfo {pages} {11} (\bibinfo {year} {1939})}\BibitemShut {NoStop}%
\bibitem [{\citenamefont {Meitner}\ and\ \citenamefont
  {Frisch}(1939)}]{MEITNER1939}%
  \BibitemOpen
  \bibfield  {author} {\bibinfo {author} {\bibfnamefont {L.}~\bibnamefont
  {Meitner}}\ and\ \bibinfo {author} {\bibfnamefont {O.~R.}\ \bibnamefont
  {Frisch}},\ }\href {\doibase 10.1038/143471a0} {\bibfield  {journal}
  {\bibinfo  {journal} {Nature}\ }\textbf {\bibinfo {volume} {143}},\ \bibinfo
  {pages} {471} (\bibinfo {year} {1939})}\BibitemShut {NoStop}%
\bibitem [{\citenamefont {Goriely}(2015)}]{Goriely2015}%
  \BibitemOpen
  \bibfield  {author} {\bibinfo {author} {\bibfnamefont {S.}~\bibnamefont
  {Goriely}},\ }\href {\doibase https://doi.org/10.1140/epja/i2015-15022-3}
  {\bibfield  {journal} {\bibinfo  {journal} {The European Physical Journal A}\
  }\textbf {\bibinfo {volume} {51}},\ \bibinfo {pages} {22} (\bibinfo {year}
  {2015})}\BibitemShut {NoStop}%
\bibitem [{\citenamefont {Vassh}\ \emph {et~al.}(2019)\citenamefont {Vassh},
  \citenamefont {Vogt}, \citenamefont {Surman}, \citenamefont {Randrup},
  \citenamefont {Sprouse}, \citenamefont {Mumpower}, \citenamefont {Jaffke},
  \citenamefont {Shaw}, \citenamefont {Holmbeck}, \citenamefont {Zhu},\ and\
  \citenamefont {McLaughlin}}]{Vassh2019}%
  \BibitemOpen
  \bibfield  {author} {\bibinfo {author} {\bibfnamefont {N.}~\bibnamefont
  {Vassh}}, \bibinfo {author} {\bibfnamefont {R.}~\bibnamefont {Vogt}},
  \bibinfo {author} {\bibfnamefont {R.}~\bibnamefont {Surman}}, \bibinfo
  {author} {\bibfnamefont {J.}~\bibnamefont {Randrup}}, \bibinfo {author}
  {\bibfnamefont {T.~M.}\ \bibnamefont {Sprouse}}, \bibinfo {author}
  {\bibfnamefont {M.~R.}\ \bibnamefont {Mumpower}}, \bibinfo {author}
  {\bibfnamefont {P.}~\bibnamefont {Jaffke}}, \bibinfo {author} {\bibfnamefont
  {D.}~\bibnamefont {Shaw}}, \bibinfo {author} {\bibfnamefont {E.~M.}\
  \bibnamefont {Holmbeck}}, \bibinfo {author} {\bibfnamefont {Y.}~\bibnamefont
  {Zhu}}, \ and\ \bibinfo {author} {\bibfnamefont {G.~C.}\ \bibnamefont
  {McLaughlin}},\ }\href {\doibase 10.1088/1361-6471/ab0bea} {\bibfield
  {journal} {\bibinfo  {journal} {Journal of Physics G: Nuclear and Particle
  Physics}\ }\textbf {\bibinfo {volume} {46}},\ \bibinfo {pages} {065202}
  (\bibinfo {year} {2019})}\BibitemShut {NoStop}%
\bibitem [{\citenamefont {Mumpower}\ \emph {et~al.}(2020)\citenamefont
  {Mumpower}, \citenamefont {Jaffke}, \citenamefont {Verriere},\ and\
  \citenamefont {Randrup}}]{Mumpower2020}%
  \BibitemOpen
  \bibfield  {author} {\bibinfo {author} {\bibfnamefont {M.~R.}\ \bibnamefont
  {Mumpower}}, \bibinfo {author} {\bibfnamefont {P.}~\bibnamefont {Jaffke}},
  \bibinfo {author} {\bibfnamefont {M.}~\bibnamefont {Verriere}}, \ and\
  \bibinfo {author} {\bibfnamefont {J.}~\bibnamefont {Randrup}},\ }\href
  {\doibase 10.1103/PhysRevC.101.054607} {\bibfield  {journal} {\bibinfo
  {journal} {Phys. Rev. C}\ }\textbf {\bibinfo {volume} {101}},\ \bibinfo
  {pages} {054607} (\bibinfo {year} {2020})}\BibitemShut {NoStop}%
\bibitem [{\citenamefont {Vassh}\ \emph {et~al.}(2020)\citenamefont {Vassh},
  \citenamefont {Mumpower}, \citenamefont {McLaughlin}, \citenamefont
  {Sprouse},\ and\ \citenamefont {Surman}}]{Vassh2020}%
  \BibitemOpen
  \bibfield  {author} {\bibinfo {author} {\bibfnamefont {N.}~\bibnamefont
  {Vassh}}, \bibinfo {author} {\bibfnamefont {M.~R.}\ \bibnamefont {Mumpower}},
  \bibinfo {author} {\bibfnamefont {G.~C.}\ \bibnamefont {McLaughlin}},
  \bibinfo {author} {\bibfnamefont {T.~M.}\ \bibnamefont {Sprouse}}, \ and\
  \bibinfo {author} {\bibfnamefont {R.}~\bibnamefont {Surman}},\ }\href
  {\doibase https://doi.org/10.3847/1538-4357/ab91a9} {\bibfield  {journal}
  {\bibinfo  {journal} {The Astrophysical Journal}\ }\textbf {\bibinfo {volume}
  {896}},\ \bibinfo {pages} {28} (\bibinfo {year} {2020})}\BibitemShut
  {NoStop}%
\bibitem [{\citenamefont {Wang}\ \emph {et~al.}(2020)\citenamefont {Wang},
  \citenamefont {Vassh}, \citenamefont {Sprouse}, \citenamefont {Mumpower},
  \citenamefont {Vogt}, \citenamefont {Randrup},\ and\ \citenamefont
  {Surman}}]{Wang2020}%
  \BibitemOpen
  \bibfield  {author} {\bibinfo {author} {\bibfnamefont {X.}~\bibnamefont
  {Wang}}, \bibinfo {author} {\bibfnamefont {N.}~\bibnamefont {Vassh}},
  \bibinfo {author} {\bibfnamefont {T.}~\bibnamefont {Sprouse}}, \bibinfo
  {author} {\bibfnamefont {M.}~\bibnamefont {Mumpower}}, \bibinfo {author}
  {\bibfnamefont {R.}~\bibnamefont {Vogt}}, \bibinfo {author} {\bibfnamefont
  {J.}~\bibnamefont {Randrup}}, \ and\ \bibinfo {author} {\bibfnamefont
  {R.}~\bibnamefont {Surman}},\ }\href {\doibase 10.3847/2041-8213/abbe18}
  {\bibfield  {journal} {\bibinfo  {journal} {The Astrophysical Journal}\
  }\textbf {\bibinfo {volume} {903}},\ \bibinfo {pages} {L3} (\bibinfo {year}
  {2020})}\BibitemShut {NoStop}%
\bibitem [{\citenamefont {Zagrebaev}\ \emph {et~al.}(2001)\citenamefont
  {Zagrebaev}, \citenamefont {Aritomo}, \citenamefont {Itkis}, \citenamefont
  {Oganessian},\ and\ \citenamefont {Ohta}}]{Zagrebaev2001}%
  \BibitemOpen
  \bibfield  {author} {\bibinfo {author} {\bibfnamefont {V.~I.}\ \bibnamefont
  {Zagrebaev}}, \bibinfo {author} {\bibfnamefont {Y.}~\bibnamefont {Aritomo}},
  \bibinfo {author} {\bibfnamefont {M.~G.}\ \bibnamefont {Itkis}}, \bibinfo
  {author} {\bibfnamefont {Y.~T.}\ \bibnamefont {Oganessian}}, \ and\ \bibinfo
  {author} {\bibfnamefont {M.}~\bibnamefont {Ohta}},\ }\href {\doibase
  https://doi.org/10.1103/PhysRevC.65.014607} {\bibfield  {journal} {\bibinfo
  {journal} {Phys. Rev. C}\ }\textbf {\bibinfo {volume} {65}},\ \bibinfo
  {pages} {014607} (\bibinfo {year} {2001})}\BibitemShut {NoStop}%
\bibitem [{\citenamefont {Itkis}\ \emph {et~al.}(2015)\citenamefont {Itkis},
  \citenamefont {Vardaci}, \citenamefont {Itkis}, \citenamefont {Knyazheva},\
  and\ \citenamefont {Kozulin}}]{Itkis2015}%
  \BibitemOpen
  \bibfield  {author} {\bibinfo {author} {\bibfnamefont {M.}~\bibnamefont
  {Itkis}}, \bibinfo {author} {\bibfnamefont {E.}~\bibnamefont {Vardaci}},
  \bibinfo {author} {\bibfnamefont {I.}~\bibnamefont {Itkis}}, \bibinfo
  {author} {\bibfnamefont {G.}~\bibnamefont {Knyazheva}}, \ and\ \bibinfo
  {author} {\bibfnamefont {E.}~\bibnamefont {Kozulin}},\ }\href {\doibase
  https://doi.org/10.1016/j.nuclphysa.2015.09.007} {\bibfield  {journal}
  {\bibinfo  {journal} {Nuclear Physics A}\ }\textbf {\bibinfo {volume}
  {944}},\ \bibinfo {pages} {204} (\bibinfo {year} {2015})},\ \bibinfo {note}
  {special Issue on Superheavy Elements}\BibitemShut {NoStop}%
\bibitem [{\citenamefont {Rimpault}\ \emph {et~al.}(2012)\citenamefont
  {Rimpault}, \citenamefont {Bernard}, \citenamefont {Blanchet}, \citenamefont
  {Vaglio-Gaudard}, \citenamefont {Ravaux},\ and\ \citenamefont
  {Santamarina}}]{Rimpault2012}%
  \BibitemOpen
  \bibfield  {author} {\bibinfo {author} {\bibfnamefont {G.}~\bibnamefont
  {Rimpault}}, \bibinfo {author} {\bibfnamefont {D.}~\bibnamefont {Bernard}},
  \bibinfo {author} {\bibfnamefont {D.}~\bibnamefont {Blanchet}}, \bibinfo
  {author} {\bibfnamefont {C.}~\bibnamefont {Vaglio-Gaudard}}, \bibinfo
  {author} {\bibfnamefont {S.}~\bibnamefont {Ravaux}}, \ and\ \bibinfo {author}
  {\bibfnamefont {A.}~\bibnamefont {Santamarina}},\ }\href {\doibase
  https://doi.org/10.1016/j.phpro.2012.04.002} {\bibfield  {journal} {\bibinfo
  {journal} {Physics Procedia}\ }\textbf {\bibinfo {volume} {31}},\ \bibinfo
  {pages} {3} (\bibinfo {year} {2012})}\BibitemShut {NoStop}%
\bibitem [{\citenamefont {Talou}\ \emph {et~al.}(2021)\citenamefont {Talou},
  \citenamefont {Stetcu}, \citenamefont {Jaffke}, \citenamefont {Rising},
  \citenamefont {Lovell},\ and\ \citenamefont {Kawano}}]{Talou2021}%
  \BibitemOpen
  \bibfield  {author} {\bibinfo {author} {\bibfnamefont {P.}~\bibnamefont
  {Talou}}, \bibinfo {author} {\bibfnamefont {I.}~\bibnamefont {Stetcu}},
  \bibinfo {author} {\bibfnamefont {P.}~\bibnamefont {Jaffke}}, \bibinfo
  {author} {\bibfnamefont {M.~E.}\ \bibnamefont {Rising}}, \bibinfo {author}
  {\bibfnamefont {A.~E.}\ \bibnamefont {Lovell}}, \ and\ \bibinfo {author}
  {\bibfnamefont {T.}~\bibnamefont {Kawano}},\ }\href {\doibase
  https://doi.org/10.1016/j.cpc.2021.108087} {\bibfield  {journal} {\bibinfo
  {journal} {Computer Physics Communications}\ }\textbf {\bibinfo {volume}
  {269}} (\bibinfo {year} {2021}),\
  https://doi.org/10.1016/j.cpc.2021.108087}\BibitemShut {NoStop}%
\bibitem [{\citenamefont {Litaize}\ \emph {et~al.}(2015)\citenamefont
  {Litaize}, \citenamefont {Serot},\ and\ \citenamefont {Berge}}]{Litaize2015}%
  \BibitemOpen
  \bibfield  {author} {\bibinfo {author} {\bibfnamefont {O.}~\bibnamefont
  {Litaize}}, \bibinfo {author} {\bibfnamefont {O.}~\bibnamefont {Serot}}, \
  and\ \bibinfo {author} {\bibfnamefont {L.}~\bibnamefont {Berge}},\ }\href
  {\doibase 10.1140/epja/i2015-15177-9} {\bibfield  {journal} {\bibinfo
  {journal} {The European Physical Journal A}\ }\textbf {\bibinfo {volume}
  {51}},\ \bibinfo {pages} {177} (\bibinfo {year} {2015})}\BibitemShut
  {NoStop}%
\bibitem [{\citenamefont {Randrup}\ and\ \citenamefont
  {Vogt}(2009)}]{Vogt2009}%
  \BibitemOpen
  \bibfield  {author} {\bibinfo {author} {\bibfnamefont {J.}~\bibnamefont
  {Randrup}}\ and\ \bibinfo {author} {\bibfnamefont {R.}~\bibnamefont {Vogt}},\
  }\href {\doibase 10.1103/PhysRevC.80.024601} {\bibfield  {journal} {\bibinfo
  {journal} {Phys. Rev. C}\ }\textbf {\bibinfo {volume} {80}},\ \bibinfo
  {pages} {024601} (\bibinfo {year} {2009})}\BibitemShut {NoStop}%
\bibitem [{\citenamefont {Vogt}\ and\ \citenamefont
  {Randrup}(2021)}]{Vogt2021}%
  \BibitemOpen
  \bibfield  {author} {\bibinfo {author} {\bibfnamefont {R.}~\bibnamefont
  {Vogt}}\ and\ \bibinfo {author} {\bibfnamefont {J.}~\bibnamefont {Randrup}},\
  }\href {\doibase 10.1103/PhysRevC.103.014610} {\bibfield  {journal} {\bibinfo
   {journal} {Phys. Rev. C}\ }\textbf {\bibinfo {volume} {103}},\ \bibinfo
  {pages} {014610} (\bibinfo {year} {2021})}\BibitemShut {NoStop}%
\bibitem [{\citenamefont {Wilson}\ \emph {et~al.}(2021)\citenamefont {Wilson},
  \citenamefont {Thisse}, \citenamefont {Lebois}, \citenamefont
  {Jovan{\v{c}}evi{\'{c}}}, \citenamefont {Gjestvang}, \citenamefont {Canavan},
  \citenamefont {Rudigier}, \citenamefont {{\'{E}}tasse}, \citenamefont
  {Gerst}, \citenamefont {Gaudefroy}, \citenamefont {Adamska}, \citenamefont
  {Adsley}, \citenamefont {Algora}, \citenamefont {Babo}, \citenamefont
  {Belvedere}, \citenamefont {Benito}, \citenamefont {Benzoni}, \citenamefont
  {Blazhev}, \citenamefont {Boso}, \citenamefont {Bottoni}, \citenamefont
  {Bunce}, \citenamefont {Chakma}, \citenamefont {Cieplicka-Ory{\'{n}}czak},
  \citenamefont {Courtin}, \citenamefont {Cort{\'{e}}s}, \citenamefont
  {Davies}, \citenamefont {Delafosse}, \citenamefont {Fallot}, \citenamefont
  {Fornal}, \citenamefont {Fraile}, \citenamefont {Gottardo}, \citenamefont
  {Guadilla}, \citenamefont {H{\"{a}}fner}, \citenamefont {Hauschild},
  \citenamefont {Heine}, \citenamefont {Henrich}, \citenamefont {Homm},
  \citenamefont {Ibrahim}, \citenamefont {Iskra}, \citenamefont {Ivanov},
  \citenamefont {Jazrawi}, \citenamefont {Korgul}, \citenamefont {Koseoglou},
  \citenamefont {Kr{\"{o}}ll}, \citenamefont {Kurtukian-Nieto}, \citenamefont
  {{Le Meur}}, \citenamefont {Leoni}, \citenamefont {Ljungvall}, \citenamefont
  {Lopez-Martens}, \citenamefont {Lozeva}, \citenamefont {Matea}, \citenamefont
  {Miernik}, \citenamefont {Nemer}, \citenamefont {Oberstedt}, \citenamefont
  {Paulsen}, \citenamefont {Piersa}, \citenamefont {Popovitch}, \citenamefont
  {Porzio}, \citenamefont {Qi}, \citenamefont {Ralet}, \citenamefont {Regan},
  \citenamefont {Rezynkina}, \citenamefont {S{\'{a}}nchez-Tembleque},
  \citenamefont {Siem}, \citenamefont {Schmitt}, \citenamefont
  {S{\"{o}}derstr{\"{o}}m}, \citenamefont {S{\"{u}}rder}, \citenamefont
  {Tocabens}, \citenamefont {Vedia}, \citenamefont {Verney}, \citenamefont
  {Warr}, \citenamefont {Wasilewska}, \citenamefont {Wiederhold}, \citenamefont
  {Yavahchova}, \citenamefont {Zeiser},\ and\ \citenamefont
  {Ziliani}}]{Wilson2021}%
  \BibitemOpen
  \bibfield  {author} {\bibinfo {author} {\bibfnamefont {J.~N.}\ \bibnamefont
  {Wilson}}, \bibinfo {author} {\bibfnamefont {D.}~\bibnamefont {Thisse}},
  \bibinfo {author} {\bibfnamefont {M.}~\bibnamefont {Lebois}}, \bibinfo
  {author} {\bibfnamefont {N.}~\bibnamefont {Jovan{\v{c}}evi{\'{c}}}}, \bibinfo
  {author} {\bibfnamefont {D.}~\bibnamefont {Gjestvang}}, \bibinfo {author}
  {\bibfnamefont {R.}~\bibnamefont {Canavan}}, \bibinfo {author} {\bibfnamefont
  {M.}~\bibnamefont {Rudigier}}, \bibinfo {author} {\bibfnamefont
  {D.}~\bibnamefont {{\'{E}}tasse}}, \bibinfo {author} {\bibfnamefont {R.-B.}\
  \bibnamefont {Gerst}}, \bibinfo {author} {\bibfnamefont {L.}~\bibnamefont
  {Gaudefroy}}, \bibinfo {author} {\bibfnamefont {E.}~\bibnamefont {Adamska}},
  \bibinfo {author} {\bibfnamefont {P.}~\bibnamefont {Adsley}}, \bibinfo
  {author} {\bibfnamefont {A.}~\bibnamefont {Algora}}, \bibinfo {author}
  {\bibfnamefont {M.}~\bibnamefont {Babo}}, \bibinfo {author} {\bibfnamefont
  {K.}~\bibnamefont {Belvedere}}, \bibinfo {author} {\bibfnamefont
  {J.}~\bibnamefont {Benito}}, \bibinfo {author} {\bibfnamefont
  {G.}~\bibnamefont {Benzoni}}, \bibinfo {author} {\bibfnamefont
  {A.}~\bibnamefont {Blazhev}}, \bibinfo {author} {\bibfnamefont
  {A.}~\bibnamefont {Boso}}, \bibinfo {author} {\bibfnamefont {S.}~\bibnamefont
  {Bottoni}}, \bibinfo {author} {\bibfnamefont {M.}~\bibnamefont {Bunce}},
  \bibinfo {author} {\bibfnamefont {R.}~\bibnamefont {Chakma}}, \bibinfo
  {author} {\bibfnamefont {N.}~\bibnamefont {Cieplicka-Ory{\'{n}}czak}},
  \bibinfo {author} {\bibfnamefont {S.}~\bibnamefont {Courtin}}, \bibinfo
  {author} {\bibfnamefont {M.~L.}\ \bibnamefont {Cort{\'{e}}s}}, \bibinfo
  {author} {\bibfnamefont {P.}~\bibnamefont {Davies}}, \bibinfo {author}
  {\bibfnamefont {C.}~\bibnamefont {Delafosse}}, \bibinfo {author}
  {\bibfnamefont {M.}~\bibnamefont {Fallot}}, \bibinfo {author} {\bibfnamefont
  {B.}~\bibnamefont {Fornal}}, \bibinfo {author} {\bibfnamefont
  {L.}~\bibnamefont {Fraile}}, \bibinfo {author} {\bibfnamefont
  {A.}~\bibnamefont {Gottardo}}, \bibinfo {author} {\bibfnamefont
  {V.}~\bibnamefont {Guadilla}}, \bibinfo {author} {\bibfnamefont
  {G.}~\bibnamefont {H{\"{a}}fner}}, \bibinfo {author} {\bibfnamefont
  {K.}~\bibnamefont {Hauschild}}, \bibinfo {author} {\bibfnamefont
  {M.}~\bibnamefont {Heine}}, \bibinfo {author} {\bibfnamefont
  {C.}~\bibnamefont {Henrich}}, \bibinfo {author} {\bibfnamefont
  {I.}~\bibnamefont {Homm}}, \bibinfo {author} {\bibfnamefont {F.}~\bibnamefont
  {Ibrahim}}, \bibinfo {author} {\bibfnamefont {{\L}.~W.}\ \bibnamefont
  {Iskra}}, \bibinfo {author} {\bibfnamefont {P.}~\bibnamefont {Ivanov}},
  \bibinfo {author} {\bibfnamefont {S.}~\bibnamefont {Jazrawi}}, \bibinfo
  {author} {\bibfnamefont {A.}~\bibnamefont {Korgul}}, \bibinfo {author}
  {\bibfnamefont {P.}~\bibnamefont {Koseoglou}}, \bibinfo {author}
  {\bibfnamefont {T.}~\bibnamefont {Kr{\"{o}}ll}}, \bibinfo {author}
  {\bibfnamefont {T.}~\bibnamefont {Kurtukian-Nieto}}, \bibinfo {author}
  {\bibfnamefont {L.}~\bibnamefont {{Le Meur}}}, \bibinfo {author}
  {\bibfnamefont {S.}~\bibnamefont {Leoni}}, \bibinfo {author} {\bibfnamefont
  {J.}~\bibnamefont {Ljungvall}}, \bibinfo {author} {\bibfnamefont
  {A.}~\bibnamefont {Lopez-Martens}}, \bibinfo {author} {\bibfnamefont
  {R.}~\bibnamefont {Lozeva}}, \bibinfo {author} {\bibfnamefont
  {I.}~\bibnamefont {Matea}}, \bibinfo {author} {\bibfnamefont
  {K.}~\bibnamefont {Miernik}}, \bibinfo {author} {\bibfnamefont
  {J.}~\bibnamefont {Nemer}}, \bibinfo {author} {\bibfnamefont
  {S.}~\bibnamefont {Oberstedt}}, \bibinfo {author} {\bibfnamefont
  {W.}~\bibnamefont {Paulsen}}, \bibinfo {author} {\bibfnamefont
  {M.}~\bibnamefont {Piersa}}, \bibinfo {author} {\bibfnamefont
  {Y.}~\bibnamefont {Popovitch}}, \bibinfo {author} {\bibfnamefont
  {C.}~\bibnamefont {Porzio}}, \bibinfo {author} {\bibfnamefont
  {L.}~\bibnamefont {Qi}}, \bibinfo {author} {\bibfnamefont {D.}~\bibnamefont
  {Ralet}}, \bibinfo {author} {\bibfnamefont {P.~H.}\ \bibnamefont {Regan}},
  \bibinfo {author} {\bibfnamefont {K.}~\bibnamefont {Rezynkina}}, \bibinfo
  {author} {\bibfnamefont {V.}~\bibnamefont {S{\'{a}}nchez-Tembleque}},
  \bibinfo {author} {\bibfnamefont {S.}~\bibnamefont {Siem}}, \bibinfo {author}
  {\bibfnamefont {C.}~\bibnamefont {Schmitt}}, \bibinfo {author} {\bibfnamefont
  {P.-A.}\ \bibnamefont {S{\"{o}}derstr{\"{o}}m}}, \bibinfo {author}
  {\bibfnamefont {C.}~\bibnamefont {S{\"{u}}rder}}, \bibinfo {author}
  {\bibfnamefont {G.}~\bibnamefont {Tocabens}}, \bibinfo {author}
  {\bibfnamefont {V.}~\bibnamefont {Vedia}}, \bibinfo {author} {\bibfnamefont
  {D.}~\bibnamefont {Verney}}, \bibinfo {author} {\bibfnamefont
  {N.}~\bibnamefont {Warr}}, \bibinfo {author} {\bibfnamefont {B.}~\bibnamefont
  {Wasilewska}}, \bibinfo {author} {\bibfnamefont {J.}~\bibnamefont
  {Wiederhold}}, \bibinfo {author} {\bibfnamefont {M.}~\bibnamefont
  {Yavahchova}}, \bibinfo {author} {\bibfnamefont {F.}~\bibnamefont {Zeiser}},
  \ and\ \bibinfo {author} {\bibfnamefont {S.}~\bibnamefont {Ziliani}},\ }\href
  {\doibase 10.1038/s41586-021-03304-w} {\bibfield  {journal} {\bibinfo
  {journal} {Nature}\ }\textbf {\bibinfo {volume} {590}},\ \bibinfo {pages}
  {566} (\bibinfo {year} {2021})}\BibitemShut {NoStop}%
\bibitem [{\citenamefont {Bulgac}\ \emph {et~al.}(2021)\citenamefont {Bulgac},
  \citenamefont {Abdurrahman}, \citenamefont {Jin}, \citenamefont {Godbey},
  \citenamefont {Schunck},\ and\ \citenamefont
  {Stetcu}}]{Bulgac_FFIntrinSpins2021}%
  \BibitemOpen
  \bibfield  {author} {\bibinfo {author} {\bibfnamefont {A.}~\bibnamefont
  {Bulgac}}, \bibinfo {author} {\bibfnamefont {I.}~\bibnamefont {Abdurrahman}},
  \bibinfo {author} {\bibfnamefont {S.}~\bibnamefont {Jin}}, \bibinfo {author}
  {\bibfnamefont {K.}~\bibnamefont {Godbey}}, \bibinfo {author} {\bibfnamefont
  {N.}~\bibnamefont {Schunck}}, \ and\ \bibinfo {author} {\bibfnamefont
  {I.}~\bibnamefont {Stetcu}},\ }\href {\doibase
  https://doi.org/10.1103/PhysRevLett.126.142502} {\bibfield  {journal}
  {\bibinfo  {journal} {Phys. Rev. Lett.}\ }\textbf {\bibinfo {volume} {126}},\
  \bibinfo {pages} {142502} (\bibinfo {year} {2021})}\BibitemShut {NoStop}%
\bibitem [{\citenamefont {Randrup}\ and\ \citenamefont
  {Vogt}(2021)}]{RandrupVogt_GenOfFragAngMom2021}%
  \BibitemOpen
  \bibfield  {author} {\bibinfo {author} {\bibfnamefont {J.}~\bibnamefont
  {Randrup}}\ and\ \bibinfo {author} {\bibfnamefont {R.}~\bibnamefont {Vogt}},\
  }\href {\doibase https://doi.org/10.1103/PhysRevLett.127.062502} {\bibfield
  {journal} {\bibinfo  {journal} {Phys. Rev. Lett.}\ }\textbf {\bibinfo
  {volume} {127}},\ \bibinfo {pages} {062502} (\bibinfo {year}
  {2021})}\BibitemShut {NoStop}%
\bibitem [{\citenamefont {Marevi\ifmmode~\acute{c}\else \'{c}\fi{}}\ \emph
  {et~al.}(2021)\citenamefont {Marevi\ifmmode~\acute{c}\else \'{c}\fi{}},
  \citenamefont {Schunck}, \citenamefont {Randrup},\ and\ \citenamefont
  {Vogt}}]{Marevic2021}%
  \BibitemOpen
  \bibfield  {author} {\bibinfo {author} {\bibfnamefont {P.}~\bibnamefont
  {Marevi\ifmmode~\acute{c}\else \'{c}\fi{}}}, \bibinfo {author} {\bibfnamefont
  {N.}~\bibnamefont {Schunck}}, \bibinfo {author} {\bibfnamefont
  {J.}~\bibnamefont {Randrup}}, \ and\ \bibinfo {author} {\bibfnamefont
  {R.}~\bibnamefont {Vogt}},\ }\href {\doibase 10.1103/PhysRevC.104.L021601}
  {\bibfield  {journal} {\bibinfo  {journal} {Phys. Rev. C}\ }\textbf {\bibinfo
  {volume} {104}},\ \bibinfo {pages} {L021601} (\bibinfo {year}
  {2021})}\BibitemShut {NoStop}%
\bibitem [{\citenamefont {Stetcu}\ \emph {et~al.}(2021)\citenamefont {Stetcu},
  \citenamefont {Lovell}, \citenamefont {Talou}, \citenamefont {Kawano},
  \citenamefont {Marin}, \citenamefont {Pozzi},\ and\ \citenamefont
  {Bulgac}}]{Stetcu2021}%
  \BibitemOpen
  \bibfield  {author} {\bibinfo {author} {\bibfnamefont {I.}~\bibnamefont
  {Stetcu}}, \bibinfo {author} {\bibfnamefont {A.~E.}\ \bibnamefont {Lovell}},
  \bibinfo {author} {\bibfnamefont {P.}~\bibnamefont {Talou}}, \bibinfo
  {author} {\bibfnamefont {T.}~\bibnamefont {Kawano}}, \bibinfo {author}
  {\bibfnamefont {S.}~\bibnamefont {Marin}}, \bibinfo {author} {\bibfnamefont
  {S.~A.}\ \bibnamefont {Pozzi}}, \ and\ \bibinfo {author} {\bibfnamefont
  {A.}~\bibnamefont {Bulgac}},\ }\href {\doibase
  https://doi.org/10.1103/PhysRevLett.127.222502} {\bibfield  {journal}
  {\bibinfo  {journal} {Phys. Rev. Lett.}\ }\textbf {\bibinfo {volume} {127}},\
  \bibinfo {pages} {222502} (\bibinfo {year} {2021})}\BibitemShut {NoStop}%
\bibitem [{\citenamefont {Gönnenwein}(2014)}]{Gonnenwein2014}%
  \BibitemOpen
  \bibfield  {author} {\bibinfo {author} {\bibfnamefont {F.}~\bibnamefont
  {Gönnenwein}},\ }in\ \href@noop {} {\emph {\bibinfo {booktitle} {Fission
  Experiments and Theoretical Advances}}}\ (\bibinfo {year} {2014})\BibitemShut
  {NoStop}%
\bibitem [{\citenamefont {Wilhelmy}\ \emph {et~al.}(1972)\citenamefont
  {Wilhelmy}, \citenamefont {Cheifetz}, \citenamefont {Jared}, \citenamefont
  {Thompson}, \citenamefont {Bowman},\ and\ \citenamefont
  {Rasmussen}}]{Wilhelmy1972}%
  \BibitemOpen
  \bibfield  {author} {\bibinfo {author} {\bibfnamefont {J.~B.}\ \bibnamefont
  {Wilhelmy}}, \bibinfo {author} {\bibfnamefont {E.}~\bibnamefont {Cheifetz}},
  \bibinfo {author} {\bibfnamefont {R.~C.}\ \bibnamefont {Jared}}, \bibinfo
  {author} {\bibfnamefont {S.~G.}\ \bibnamefont {Thompson}}, \bibinfo {author}
  {\bibfnamefont {H.~R.}\ \bibnamefont {Bowman}}, \ and\ \bibinfo {author}
  {\bibfnamefont {J.~O.}\ \bibnamefont {Rasmussen}},\ }\href {\doibase
  https://doi.org/10.1103/PhysRevC.5.2041} {\bibfield  {journal} {\bibinfo
  {journal} {Phys. Rev. C}\ }\textbf {\bibinfo {volume} {5}},\ \bibinfo {pages}
  {2041} (\bibinfo {year} {1972})}\BibitemShut {NoStop}%
\bibitem [{\citenamefont {Moretto}\ \emph {et~al.}(1989)\citenamefont
  {Moretto}, \citenamefont {Peaslee},\ and\ \citenamefont
  {Wozniak}}]{Moretto1989}%
  \BibitemOpen
  \bibfield  {author} {\bibinfo {author} {\bibfnamefont {L.~G.}\ \bibnamefont
  {Moretto}}, \bibinfo {author} {\bibfnamefont {G.~F.}\ \bibnamefont
  {Peaslee}}, \ and\ \bibinfo {author} {\bibfnamefont {G.~J.}\ \bibnamefont
  {Wozniak}},\ }\href {\doibase https://doi.org/10.1016/0375-9474(89)90682-9}
  {\bibfield  {journal} {\bibinfo  {journal} {Nuclear Physics A}\ }\textbf
  {\bibinfo {volume} {502}},\ \bibinfo {pages} {453} (\bibinfo {year}
  {1989})}\BibitemShut {NoStop}%
\bibitem [{\citenamefont {Fr{\'e}haut}\ \emph {et~al.}(1983)\citenamefont
  {Fr{\'e}haut}, \citenamefont {Bertin},\ and\ \citenamefont
  {Bois}}]{Frehaut1983}%
  \BibitemOpen
  \bibfield  {author} {\bibinfo {author} {\bibfnamefont {J.}~\bibnamefont
  {Fr{\'e}haut}}, \bibinfo {author} {\bibfnamefont {A.}~\bibnamefont {Bertin}},
  \ and\ \bibinfo {author} {\bibfnamefont {R.}~\bibnamefont {Bois}},\ }in\
  \href@noop {} {\emph {\bibinfo {booktitle} {Nuclear Data for Science and
  Technology}}},\ \bibinfo {editor} {edited by\ \bibinfo {editor}
  {\bibfnamefont {K.~H.}\ \bibnamefont {B{\"o}ckhoff}}}\ (\bibinfo  {publisher}
  {Springer Netherlands},\ \bibinfo {address} {Dordrecht},\ \bibinfo {year}
  {1983})\ pp.\ \bibinfo {pages} {78--81}\BibitemShut {NoStop}%
\bibitem [{\citenamefont {Frehaut}(1989)}]{Frehaut1989}%
  \BibitemOpen
  \bibfield  {author} {\bibinfo {author} {\bibfnamefont {J.}~\bibnamefont
  {Frehaut}},\ }\href
  {http://inis.iaea.org/search/search.aspx?orig_q=RN:20083133} {\emph {\bibinfo
  {title} {Neutron gamma competition in fast fission}}},\ \bibinfo {type}
  {Tech. Rep.}\ (\bibinfo {address} {International Atomic Energy Agency
  (IAEA)},\ \bibinfo {year} {1989})\ \bibinfo {note}
  {iNDC(NDS)--220}\BibitemShut {NoStop}%
\bibitem [{\citenamefont {Qi}\ \emph {et~al.}(2018)\citenamefont {Qi},
  \citenamefont {Lebois}, \citenamefont {Wilson}, \citenamefont {Chatillon},
  \citenamefont {Courtin}, \citenamefont {Fruet}, \citenamefont {Georgiev},
  \citenamefont {Jenkins}, \citenamefont {Laurent}, \citenamefont {Le~Meur},
  \citenamefont {Maj}, \citenamefont {Marini}, \citenamefont {Matea},
  \citenamefont {Morris}, \citenamefont {Nanal}, \citenamefont {Napiorkowski},
  \citenamefont {Oberstedt}, \citenamefont {Oberstedt}, \citenamefont
  {Schmitt}, \citenamefont {Serot}, \citenamefont {Stanoiu},\ and\
  \citenamefont {Wasilewska}}]{Qi2018}%
  \BibitemOpen
  \bibfield  {author} {\bibinfo {author} {\bibfnamefont {L.}~\bibnamefont
  {Qi}}, \bibinfo {author} {\bibfnamefont {M.}~\bibnamefont {Lebois}}, \bibinfo
  {author} {\bibfnamefont {J.~N.}\ \bibnamefont {Wilson}}, \bibinfo {author}
  {\bibfnamefont {A.}~\bibnamefont {Chatillon}}, \bibinfo {author}
  {\bibfnamefont {S.}~\bibnamefont {Courtin}}, \bibinfo {author} {\bibfnamefont
  {G.}~\bibnamefont {Fruet}}, \bibinfo {author} {\bibfnamefont
  {G.}~\bibnamefont {Georgiev}}, \bibinfo {author} {\bibfnamefont {D.~G.}\
  \bibnamefont {Jenkins}}, \bibinfo {author} {\bibfnamefont {B.}~\bibnamefont
  {Laurent}}, \bibinfo {author} {\bibfnamefont {L.}~\bibnamefont {Le~Meur}},
  \bibinfo {author} {\bibfnamefont {A.}~\bibnamefont {Maj}}, \bibinfo {author}
  {\bibfnamefont {P.}~\bibnamefont {Marini}}, \bibinfo {author} {\bibfnamefont
  {I.}~\bibnamefont {Matea}}, \bibinfo {author} {\bibfnamefont
  {L.}~\bibnamefont {Morris}}, \bibinfo {author} {\bibfnamefont
  {V.}~\bibnamefont {Nanal}}, \bibinfo {author} {\bibfnamefont
  {P.}~\bibnamefont {Napiorkowski}}, \bibinfo {author} {\bibfnamefont
  {A.}~\bibnamefont {Oberstedt}}, \bibinfo {author} {\bibfnamefont
  {S.}~\bibnamefont {Oberstedt}}, \bibinfo {author} {\bibfnamefont
  {C.}~\bibnamefont {Schmitt}}, \bibinfo {author} {\bibfnamefont
  {O.}~\bibnamefont {Serot}}, \bibinfo {author} {\bibfnamefont
  {M.}~\bibnamefont {Stanoiu}}, \ and\ \bibinfo {author} {\bibfnamefont
  {B.}~\bibnamefont {Wasilewska}},\ }\href {\doibase
  https://doi.org/10.1103/PhysRevC.98.014612} {\bibfield  {journal} {\bibinfo
  {journal} {Phys. Rev. C}\ }\textbf {\bibinfo {volume} {98}},\ \bibinfo
  {pages} {014612} (\bibinfo {year} {2018})}\BibitemShut {NoStop}%
\bibitem [{\citenamefont {Laborie}\ \emph {et~al.}(2018)\citenamefont
  {Laborie}, \citenamefont {Billnert}, \citenamefont {B\'elier}, \citenamefont
  {Oberstedt}, \citenamefont {Oberstedt},\ and\ \citenamefont
  {Taieb}}]{Laborie2018}%
  \BibitemOpen
  \bibfield  {author} {\bibinfo {author} {\bibfnamefont {J.-M.}\ \bibnamefont
  {Laborie}}, \bibinfo {author} {\bibfnamefont {R.}~\bibnamefont {Billnert}},
  \bibinfo {author} {\bibfnamefont {G.}~\bibnamefont {B\'elier}}, \bibinfo
  {author} {\bibfnamefont {A.}~\bibnamefont {Oberstedt}}, \bibinfo {author}
  {\bibfnamefont {S.}~\bibnamefont {Oberstedt}}, \ and\ \bibinfo {author}
  {\bibfnamefont {J.}~\bibnamefont {Taieb}},\ }\href {\doibase
  https://doi.org/10.1103/PhysRevC.98.054604} {\bibfield  {journal} {\bibinfo
  {journal} {Phys. Rev. C}\ }\textbf {\bibinfo {volume} {98}},\ \bibinfo
  {pages} {054604} (\bibinfo {year} {2018})}\BibitemShut {NoStop}%
\bibitem [{\citenamefont {Oberstedt}\ \emph {et~al.}(2020)\citenamefont
  {Oberstedt}, \citenamefont {Lebois}, \citenamefont {Oberstedt}, \citenamefont
  {Qi},\ and\ \citenamefont {Wilson}}]{Oberstedt2020}%
  \BibitemOpen
  \bibfield  {author} {\bibinfo {author} {\bibfnamefont {A.}~\bibnamefont
  {Oberstedt}}, \bibinfo {author} {\bibfnamefont {M.}~\bibnamefont {Lebois}},
  \bibinfo {author} {\bibfnamefont {S.}~\bibnamefont {Oberstedt}}, \bibinfo
  {author} {\bibfnamefont {L.}~\bibnamefont {Qi}}, \ and\ \bibinfo {author}
  {\bibfnamefont {J.~N.}\ \bibnamefont {Wilson}},\ }\href {\doibase
  https://doi.org/10.1140/epja/s10050-020-00246-1} {\bibfield  {journal}
  {\bibinfo  {journal} {The European Physical Journal A}\ }\textbf {\bibinfo
  {volume} {56}},\ \bibinfo {pages} {236} (\bibinfo {year} {2020})}\BibitemShut
  {NoStop}%
\bibitem [{\citenamefont {Rose}\ \emph {et~al.}(2017)\citenamefont {Rose},
  \citenamefont {Zeiser}, \citenamefont {Wilson}, \citenamefont {Oberstedt},
  \citenamefont {Oberstedt}, \citenamefont {Siem}, \citenamefont {Tveten},
  \citenamefont {Bernstein}, \citenamefont {Bleuel}, \citenamefont {Brown},
  \citenamefont {Crespo~Campo}, \citenamefont {Giacoppo}, \citenamefont
  {G\"orgen}, \citenamefont {Guttormsen}, \citenamefont
  {Hady\ifmmode~\acute{n}\else \'{n}\fi{}ska}, \citenamefont {Hafreager},
  \citenamefont {Hagen}, \citenamefont {Klintefjord}, \citenamefont {Laplace},
  \citenamefont {Larsen}, \citenamefont {Renstr\o{}m}, \citenamefont {Sahin},
  \citenamefont {Schmitt}, \citenamefont {Tornyi},\ and\ \citenamefont
  {Wiedeking}}]{Rose2017}%
  \BibitemOpen
  \bibfield  {author} {\bibinfo {author} {\bibfnamefont {S.~J.}\ \bibnamefont
  {Rose}}, \bibinfo {author} {\bibfnamefont {F.}~\bibnamefont {Zeiser}},
  \bibinfo {author} {\bibfnamefont {J.~N.}\ \bibnamefont {Wilson}}, \bibinfo
  {author} {\bibfnamefont {A.}~\bibnamefont {Oberstedt}}, \bibinfo {author}
  {\bibfnamefont {S.}~\bibnamefont {Oberstedt}}, \bibinfo {author}
  {\bibfnamefont {S.}~\bibnamefont {Siem}}, \bibinfo {author} {\bibfnamefont
  {G.~M.}\ \bibnamefont {Tveten}}, \bibinfo {author} {\bibfnamefont {L.~A.}\
  \bibnamefont {Bernstein}}, \bibinfo {author} {\bibfnamefont {D.~L.}\
  \bibnamefont {Bleuel}}, \bibinfo {author} {\bibfnamefont {J.~A.}\
  \bibnamefont {Brown}}, \bibinfo {author} {\bibfnamefont {L.}~\bibnamefont
  {Crespo~Campo}}, \bibinfo {author} {\bibfnamefont {F.}~\bibnamefont
  {Giacoppo}}, \bibinfo {author} {\bibfnamefont {A.}~\bibnamefont {G\"orgen}},
  \bibinfo {author} {\bibfnamefont {M.}~\bibnamefont {Guttormsen}}, \bibinfo
  {author} {\bibfnamefont {K.}~\bibnamefont {Hady\ifmmode~\acute{n}\else
  \'{n}\fi{}ska}}, \bibinfo {author} {\bibfnamefont {A.}~\bibnamefont
  {Hafreager}}, \bibinfo {author} {\bibfnamefont {T.~W.}\ \bibnamefont
  {Hagen}}, \bibinfo {author} {\bibfnamefont {M.}~\bibnamefont {Klintefjord}},
  \bibinfo {author} {\bibfnamefont {T.~A.}\ \bibnamefont {Laplace}}, \bibinfo
  {author} {\bibfnamefont {A.~C.}\ \bibnamefont {Larsen}}, \bibinfo {author}
  {\bibfnamefont {T.}~\bibnamefont {Renstr\o{}m}}, \bibinfo {author}
  {\bibfnamefont {E.}~\bibnamefont {Sahin}}, \bibinfo {author} {\bibfnamefont
  {C.}~\bibnamefont {Schmitt}}, \bibinfo {author} {\bibfnamefont {T.~G.}\
  \bibnamefont {Tornyi}}, \ and\ \bibinfo {author} {\bibfnamefont
  {M.}~\bibnamefont {Wiedeking}},\ }\href {\doibase 10.1103/PhysRevC.96.014601}
  {\bibfield  {journal} {\bibinfo  {journal} {Phys. Rev. C}\ }\textbf {\bibinfo
  {volume} {96}},\ \bibinfo {pages} {014601} (\bibinfo {year}
  {2017})}\BibitemShut {NoStop}%
\bibitem [{\citenamefont {Gjestvang}\ \emph {et~al.}(2021)\citenamefont
  {Gjestvang}, \citenamefont {Siem}, \citenamefont {Zeiser}, \citenamefont
  {Randrup}, \citenamefont {Vogt}, \citenamefont {Wilson}, \citenamefont
  {Bello-Garrote}, \citenamefont {Bernstein}, \citenamefont {Bleuel},
  \citenamefont {Guttormsen}, \citenamefont {G\"orgen}, \citenamefont {Larsen},
  \citenamefont {Malatji}, \citenamefont {Matthews}, \citenamefont {Oberstedt},
  \citenamefont {Oberstedt}, \citenamefont {Tornyi}, \citenamefont {Tveten},\
  and\ \citenamefont {Voyles}}]{Gjestvang2021}%
  \BibitemOpen
  \bibfield  {author} {\bibinfo {author} {\bibfnamefont {D.}~\bibnamefont
  {Gjestvang}}, \bibinfo {author} {\bibfnamefont {S.}~\bibnamefont {Siem}},
  \bibinfo {author} {\bibfnamefont {F.}~\bibnamefont {Zeiser}}, \bibinfo
  {author} {\bibfnamefont {J.}~\bibnamefont {Randrup}}, \bibinfo {author}
  {\bibfnamefont {R.}~\bibnamefont {Vogt}}, \bibinfo {author} {\bibfnamefont
  {J.~N.}\ \bibnamefont {Wilson}}, \bibinfo {author} {\bibfnamefont
  {F.}~\bibnamefont {Bello-Garrote}}, \bibinfo {author} {\bibfnamefont {L.~A.}\
  \bibnamefont {Bernstein}}, \bibinfo {author} {\bibfnamefont {D.~L.}\
  \bibnamefont {Bleuel}}, \bibinfo {author} {\bibfnamefont {M.}~\bibnamefont
  {Guttormsen}}, \bibinfo {author} {\bibfnamefont {A.}~\bibnamefont
  {G\"orgen}}, \bibinfo {author} {\bibfnamefont {A.~C.}\ \bibnamefont
  {Larsen}}, \bibinfo {author} {\bibfnamefont {K.~L.}\ \bibnamefont {Malatji}},
  \bibinfo {author} {\bibfnamefont {E.~F.}\ \bibnamefont {Matthews}}, \bibinfo
  {author} {\bibfnamefont {A.}~\bibnamefont {Oberstedt}}, \bibinfo {author}
  {\bibfnamefont {S.}~\bibnamefont {Oberstedt}}, \bibinfo {author}
  {\bibfnamefont {T.}~\bibnamefont {Tornyi}}, \bibinfo {author} {\bibfnamefont
  {G.~M.}\ \bibnamefont {Tveten}}, \ and\ \bibinfo {author} {\bibfnamefont
  {A.~S.}\ \bibnamefont {Voyles}},\ }\href {\doibase
  https://doi.org/10.1103/PhysRevC.103.034609} {\bibfield  {journal} {\bibinfo
  {journal} {Phys. Rev. C}\ }\textbf {\bibinfo {volume} {103}},\ \bibinfo
  {pages} {034609} (\bibinfo {year} {2021})}\BibitemShut {NoStop}%
\bibitem [{\citenamefont {Brown}\ \emph {et~al.}(2018)\citenamefont {Brown},
  \citenamefont {Chadwick}, \citenamefont {Capote}, \citenamefont {Kahler},
  \citenamefont {Trkov}, \citenamefont {Herman}, \citenamefont {Sonzogni},
  \citenamefont {Danon}, \citenamefont {Carlson}, \citenamefont {Dunn},
  \citenamefont {Smith}, \citenamefont {Hale}, \citenamefont {Arbanas},
  \citenamefont {Arcilla}, \citenamefont {Bates}, \citenamefont {Beck},
  \citenamefont {Becker}, \citenamefont {Brown}, \citenamefont {Casperson},
  \citenamefont {Conlin}, \citenamefont {Cullen}, \citenamefont {Descalle},
  \citenamefont {Firestone}, \citenamefont {Gaines}, \citenamefont {Guber},
  \citenamefont {Hawari}, \citenamefont {Holmes}, \citenamefont {Johnson},
  \citenamefont {Kawano}, \citenamefont {Kiedrowski}, \citenamefont {Koning},
  \citenamefont {Kopecky}, \citenamefont {Leal}, \citenamefont {Lestone},
  \citenamefont {Lubitz}, \citenamefont {{Márquez Damián}}, \citenamefont
  {Mattoon}, \citenamefont {McCutchan}, \citenamefont {Mughabghab},
  \citenamefont {Navratil}, \citenamefont {Neudecker}, \citenamefont {Nobre},
  \citenamefont {Noguere}, \citenamefont {Paris}, \citenamefont {Pigni},
  \citenamefont {Plompen}, \citenamefont {Pritychenko}, \citenamefont
  {Pronyaev}, \citenamefont {Roubtsov}, \citenamefont {Rochman}, \citenamefont
  {Romano}, \citenamefont {Schillebeeckx}, \citenamefont {Simakov},
  \citenamefont {Sin}, \citenamefont {Sirakov}, \citenamefont {Sleaford},
  \citenamefont {Sobes}, \citenamefont {Soukhovitskii}, \citenamefont {Stetcu},
  \citenamefont {Talou}, \citenamefont {Thompson}, \citenamefont {{van der
  Marck}}, \citenamefont {Welser-Sherrill}, \citenamefont {Wiarda},
  \citenamefont {White}, \citenamefont {Wormald}, \citenamefont {Wright},
  \citenamefont {Zerkle}, \citenamefont {Žerovnik},\ and\ \citenamefont
  {Zhu}}]{ENDF_BROWN20181}%
  \BibitemOpen
  \bibfield  {author} {\bibinfo {author} {\bibfnamefont {D.}~\bibnamefont
  {Brown}}, \bibinfo {author} {\bibfnamefont {M.}~\bibnamefont {Chadwick}},
  \bibinfo {author} {\bibfnamefont {R.}~\bibnamefont {Capote}}, \bibinfo
  {author} {\bibfnamefont {A.}~\bibnamefont {Kahler}}, \bibinfo {author}
  {\bibfnamefont {A.}~\bibnamefont {Trkov}}, \bibinfo {author} {\bibfnamefont
  {M.}~\bibnamefont {Herman}}, \bibinfo {author} {\bibfnamefont
  {A.}~\bibnamefont {Sonzogni}}, \bibinfo {author} {\bibfnamefont
  {Y.}~\bibnamefont {Danon}}, \bibinfo {author} {\bibfnamefont
  {A.}~\bibnamefont {Carlson}}, \bibinfo {author} {\bibfnamefont
  {M.}~\bibnamefont {Dunn}}, \bibinfo {author} {\bibfnamefont {D.}~\bibnamefont
  {Smith}}, \bibinfo {author} {\bibfnamefont {G.}~\bibnamefont {Hale}},
  \bibinfo {author} {\bibfnamefont {G.}~\bibnamefont {Arbanas}}, \bibinfo
  {author} {\bibfnamefont {R.}~\bibnamefont {Arcilla}}, \bibinfo {author}
  {\bibfnamefont {C.}~\bibnamefont {Bates}}, \bibinfo {author} {\bibfnamefont
  {B.}~\bibnamefont {Beck}}, \bibinfo {author} {\bibfnamefont {B.}~\bibnamefont
  {Becker}}, \bibinfo {author} {\bibfnamefont {F.}~\bibnamefont {Brown}},
  \bibinfo {author} {\bibfnamefont {R.}~\bibnamefont {Casperson}}, \bibinfo
  {author} {\bibfnamefont {J.}~\bibnamefont {Conlin}}, \bibinfo {author}
  {\bibfnamefont {D.}~\bibnamefont {Cullen}}, \bibinfo {author} {\bibfnamefont
  {M.-A.}\ \bibnamefont {Descalle}}, \bibinfo {author} {\bibfnamefont
  {R.}~\bibnamefont {Firestone}}, \bibinfo {author} {\bibfnamefont
  {T.}~\bibnamefont {Gaines}}, \bibinfo {author} {\bibfnamefont
  {K.}~\bibnamefont {Guber}}, \bibinfo {author} {\bibfnamefont
  {A.}~\bibnamefont {Hawari}}, \bibinfo {author} {\bibfnamefont
  {J.}~\bibnamefont {Holmes}}, \bibinfo {author} {\bibfnamefont
  {T.}~\bibnamefont {Johnson}}, \bibinfo {author} {\bibfnamefont
  {T.}~\bibnamefont {Kawano}}, \bibinfo {author} {\bibfnamefont
  {B.}~\bibnamefont {Kiedrowski}}, \bibinfo {author} {\bibfnamefont
  {A.}~\bibnamefont {Koning}}, \bibinfo {author} {\bibfnamefont
  {S.}~\bibnamefont {Kopecky}}, \bibinfo {author} {\bibfnamefont
  {L.}~\bibnamefont {Leal}}, \bibinfo {author} {\bibfnamefont {J.}~\bibnamefont
  {Lestone}}, \bibinfo {author} {\bibfnamefont {C.}~\bibnamefont {Lubitz}},
  \bibinfo {author} {\bibfnamefont {J.}~\bibnamefont {{Márquez Damián}}},
  \bibinfo {author} {\bibfnamefont {C.}~\bibnamefont {Mattoon}}, \bibinfo
  {author} {\bibfnamefont {E.}~\bibnamefont {McCutchan}}, \bibinfo {author}
  {\bibfnamefont {S.}~\bibnamefont {Mughabghab}}, \bibinfo {author}
  {\bibfnamefont {P.}~\bibnamefont {Navratil}}, \bibinfo {author}
  {\bibfnamefont {D.}~\bibnamefont {Neudecker}}, \bibinfo {author}
  {\bibfnamefont {G.}~\bibnamefont {Nobre}}, \bibinfo {author} {\bibfnamefont
  {G.}~\bibnamefont {Noguere}}, \bibinfo {author} {\bibfnamefont
  {M.}~\bibnamefont {Paris}}, \bibinfo {author} {\bibfnamefont
  {M.}~\bibnamefont {Pigni}}, \bibinfo {author} {\bibfnamefont
  {A.}~\bibnamefont {Plompen}}, \bibinfo {author} {\bibfnamefont
  {B.}~\bibnamefont {Pritychenko}}, \bibinfo {author} {\bibfnamefont
  {V.}~\bibnamefont {Pronyaev}}, \bibinfo {author} {\bibfnamefont
  {D.}~\bibnamefont {Roubtsov}}, \bibinfo {author} {\bibfnamefont
  {D.}~\bibnamefont {Rochman}}, \bibinfo {author} {\bibfnamefont
  {P.}~\bibnamefont {Romano}}, \bibinfo {author} {\bibfnamefont
  {P.}~\bibnamefont {Schillebeeckx}}, \bibinfo {author} {\bibfnamefont
  {S.}~\bibnamefont {Simakov}}, \bibinfo {author} {\bibfnamefont
  {M.}~\bibnamefont {Sin}}, \bibinfo {author} {\bibfnamefont {I.}~\bibnamefont
  {Sirakov}}, \bibinfo {author} {\bibfnamefont {B.}~\bibnamefont {Sleaford}},
  \bibinfo {author} {\bibfnamefont {V.}~\bibnamefont {Sobes}}, \bibinfo
  {author} {\bibfnamefont {E.}~\bibnamefont {Soukhovitskii}}, \bibinfo {author}
  {\bibfnamefont {I.}~\bibnamefont {Stetcu}}, \bibinfo {author} {\bibfnamefont
  {P.}~\bibnamefont {Talou}}, \bibinfo {author} {\bibfnamefont
  {I.}~\bibnamefont {Thompson}}, \bibinfo {author} {\bibfnamefont
  {S.}~\bibnamefont {{van der Marck}}}, \bibinfo {author} {\bibfnamefont
  {L.}~\bibnamefont {Welser-Sherrill}}, \bibinfo {author} {\bibfnamefont
  {D.}~\bibnamefont {Wiarda}}, \bibinfo {author} {\bibfnamefont
  {M.}~\bibnamefont {White}}, \bibinfo {author} {\bibfnamefont
  {J.}~\bibnamefont {Wormald}}, \bibinfo {author} {\bibfnamefont
  {R.}~\bibnamefont {Wright}}, \bibinfo {author} {\bibfnamefont
  {M.}~\bibnamefont {Zerkle}}, \bibinfo {author} {\bibfnamefont
  {G.}~\bibnamefont {Žerovnik}}, \ and\ \bibinfo {author} {\bibfnamefont
  {Y.}~\bibnamefont {Zhu}},\ }\href {\doibase
  https://doi.org/10.1016/j.nds.2018.02.001} {\bibfield  {journal} {\bibinfo
  {journal} {Nuclear Data Sheets}\ }\textbf {\bibinfo {volume} {148}},\
  \bibinfo {pages} {1} (\bibinfo {year} {2018})},\ \bibinfo {note} {special
  Issue on Nuclear Reaction Data}\BibitemShut {NoStop}%
\bibitem [{\citenamefont {Kelly}\ \emph {et~al.}(2020)\citenamefont {Kelly},
  \citenamefont {Devlin}, \citenamefont {O'Donnell}, \citenamefont {Gomez},
  \citenamefont {Neudecker}, \citenamefont {Haight}, \citenamefont {Taddeucci},
  \citenamefont {Mosby}, \citenamefont {Lee}, \citenamefont {Wu}, \citenamefont
  {Henderson}, \citenamefont {Talou}, \citenamefont {Kawano}, \citenamefont
  {Lovell}, \citenamefont {White}, \citenamefont {Ullmann}, \citenamefont
  {Fotiades}, \citenamefont {Henderson},\ and\ \citenamefont
  {Buckner}}]{Kelly_PFNS_2020}%
  \BibitemOpen
  \bibfield  {author} {\bibinfo {author} {\bibfnamefont {K.~J.}\ \bibnamefont
  {Kelly}}, \bibinfo {author} {\bibfnamefont {M.}~\bibnamefont {Devlin}},
  \bibinfo {author} {\bibfnamefont {J.~M.}\ \bibnamefont {O'Donnell}}, \bibinfo
  {author} {\bibfnamefont {J.~A.}\ \bibnamefont {Gomez}}, \bibinfo {author}
  {\bibfnamefont {D.}~\bibnamefont {Neudecker}}, \bibinfo {author}
  {\bibfnamefont {R.~C.}\ \bibnamefont {Haight}}, \bibinfo {author}
  {\bibfnamefont {T.~N.}\ \bibnamefont {Taddeucci}}, \bibinfo {author}
  {\bibfnamefont {S.~M.}\ \bibnamefont {Mosby}}, \bibinfo {author}
  {\bibfnamefont {H.~Y.}\ \bibnamefont {Lee}}, \bibinfo {author} {\bibfnamefont
  {C.~Y.}\ \bibnamefont {Wu}}, \bibinfo {author} {\bibfnamefont
  {R.}~\bibnamefont {Henderson}}, \bibinfo {author} {\bibfnamefont
  {P.}~\bibnamefont {Talou}}, \bibinfo {author} {\bibfnamefont
  {T.}~\bibnamefont {Kawano}}, \bibinfo {author} {\bibfnamefont {A.~E.}\
  \bibnamefont {Lovell}}, \bibinfo {author} {\bibfnamefont {M.~C.}\
  \bibnamefont {White}}, \bibinfo {author} {\bibfnamefont {J.~L.}\ \bibnamefont
  {Ullmann}}, \bibinfo {author} {\bibfnamefont {N.}~\bibnamefont {Fotiades}},
  \bibinfo {author} {\bibfnamefont {J.}~\bibnamefont {Henderson}}, \ and\
  \bibinfo {author} {\bibfnamefont {M.~Q.}\ \bibnamefont {Buckner}},\ }\href
  {\doibase 10.1103/PhysRevC.102.034615} {\bibfield  {journal} {\bibinfo
  {journal} {Phys. Rev. C}\ }\textbf {\bibinfo {volume} {102}},\ \bibinfo
  {pages} {034615} (\bibinfo {year} {2020})}\BibitemShut {NoStop}%
\bibitem [{lan()}]{lansce}%
  \BibitemOpen
  \href@noop {} {}\bibinfo {howpublished}
  {\url{https://lansce.lanl.gov}}\BibitemShut {NoStop}%
\bibitem [{\citenamefont {Wu}\ \emph {et~al.}(2015)\citenamefont {Wu},
  \citenamefont {Henderson}, \citenamefont {Haight}, \citenamefont {Lee},
  \citenamefont {Taddeucci}, \citenamefont {Bucher}, \citenamefont {Chyzh},
  \citenamefont {Devlin}, \citenamefont {Fotiades}, \citenamefont {Kwan},
  \citenamefont {O’Donnell}, \citenamefont {Perdue},\ and\ \citenamefont
  {Ullmann}}]{ppac_2015}%
  \BibitemOpen
  \bibfield  {author} {\bibinfo {author} {\bibfnamefont {C.}~\bibnamefont
  {Wu}}, \bibinfo {author} {\bibfnamefont {R.}~\bibnamefont {Henderson}},
  \bibinfo {author} {\bibfnamefont {R.}~\bibnamefont {Haight}}, \bibinfo
  {author} {\bibfnamefont {H.}~\bibnamefont {Lee}}, \bibinfo {author}
  {\bibfnamefont {T.}~\bibnamefont {Taddeucci}}, \bibinfo {author}
  {\bibfnamefont {B.}~\bibnamefont {Bucher}}, \bibinfo {author} {\bibfnamefont
  {A.}~\bibnamefont {Chyzh}}, \bibinfo {author} {\bibfnamefont
  {M.}~\bibnamefont {Devlin}}, \bibinfo {author} {\bibfnamefont
  {N.}~\bibnamefont {Fotiades}}, \bibinfo {author} {\bibfnamefont
  {E.}~\bibnamefont {Kwan}}, \bibinfo {author} {\bibfnamefont {J.}~\bibnamefont
  {O’Donnell}}, \bibinfo {author} {\bibfnamefont {B.}~\bibnamefont {Perdue}},
  \ and\ \bibinfo {author} {\bibfnamefont {J.}~\bibnamefont {Ullmann}},\ }\href
  {\doibase https://doi.org/10.1016/j.nima.2015.05.010} {\bibfield  {journal}
  {\bibinfo  {journal} {Nuclear Instruments and Methods in Physics Research
  Section A: Accelerators, Spectrometers, Detectors and Associated Equipment}\
  }\textbf {\bibinfo {volume} {794}},\ \bibinfo {pages} {76} (\bibinfo {year}
  {2015})}\BibitemShut {NoStop}%
\bibitem [{ej3()}]{ej309}%
  \BibitemOpen
  \href@noop {} {}\bibinfo {howpublished}
  {\url{https://eljentechnology.com/products/liquid-scintillators/ej-301-ej-309}}\BibitemShut
  {NoStop}%
\bibitem [{\citenamefont {Marin}\ \emph {et~al.}(2020)\citenamefont {Marin},
  \citenamefont {Protopopescu}, \citenamefont {Vogt}, \citenamefont {Marcath},
  \citenamefont {Okar}, \citenamefont {Hua}, \citenamefont {Talou},
  \citenamefont {Schuster}, \citenamefont {Clarke},\ and\ \citenamefont
  {Pozzi}}]{Marin2020}%
  \BibitemOpen
  \bibfield  {author} {\bibinfo {author} {\bibfnamefont {S.}~\bibnamefont
  {Marin}}, \bibinfo {author} {\bibfnamefont {V.~A.}\ \bibnamefont
  {Protopopescu}}, \bibinfo {author} {\bibfnamefont {R.}~\bibnamefont {Vogt}},
  \bibinfo {author} {\bibfnamefont {M.~J.}\ \bibnamefont {Marcath}}, \bibinfo
  {author} {\bibfnamefont {S.}~\bibnamefont {Okar}}, \bibinfo {author}
  {\bibfnamefont {M.~Y.}\ \bibnamefont {Hua}}, \bibinfo {author} {\bibfnamefont
  {P.}~\bibnamefont {Talou}}, \bibinfo {author} {\bibfnamefont {P.~F.}\
  \bibnamefont {Schuster}}, \bibinfo {author} {\bibfnamefont {S.~D.}\
  \bibnamefont {Clarke}}, \ and\ \bibinfo {author} {\bibfnamefont {S.~A.}\
  \bibnamefont {Pozzi}},\ }\href {\doibase
  https://doi.org/10.1016/j.nima.2020.163907} {\bibfield  {journal} {\bibinfo
  {journal} {Nuclear Instruments and Methods in Physics Research Section A:
  Accelerators, Spectrometers, Detectors and Associated Equipment}\ }\textbf
  {\bibinfo {volume} {968}},\ \bibinfo {pages} {163907} (\bibinfo {year}
  {2020})}\BibitemShut {NoStop}%
\bibitem [{\citenamefont {Pozzi}\ \emph {et~al.}(2003)\citenamefont {Pozzi},
  \citenamefont {Padovani},\ and\ \citenamefont {Marseguerra}}]{Pozzi2003}%
  \BibitemOpen
  \bibfield  {author} {\bibinfo {author} {\bibfnamefont {S.~A.}\ \bibnamefont
  {Pozzi}}, \bibinfo {author} {\bibfnamefont {E.}~\bibnamefont {Padovani}}, \
  and\ \bibinfo {author} {\bibfnamefont {M.}~\bibnamefont {Marseguerra}},\
  }\href {\doibase https://doi.org/10.1016/j.nima.2003.06.012} {\bibfield
  {journal} {\bibinfo  {journal} {Nuclear Instruments and Methods in Physics
  Research Section A: Accelerators, Spectrometers, Detectors and Associated
  Equipment}\ }\textbf {\bibinfo {volume} {513}},\ \bibinfo {pages} {550 }
  (\bibinfo {year} {2003})}\BibitemShut {NoStop}%
\bibitem [{\citenamefont {Schmitt}(2017)}]{Schmitt2016}%
  \BibitemOpen
  \bibfield  {author} {\bibinfo {author} {\bibfnamefont {S.}~\bibnamefont
  {Schmitt}},\ }\href {\doibase 10.1051/epjconf/201713711008} {\bibfield
  {journal} {\bibinfo  {journal} {EPJ Web Conf.}\ }\textbf {\bibinfo {volume}
  {137}},\ \bibinfo {pages} {11008} (\bibinfo {year} {2017})}\BibitemShut
  {NoStop}%
\bibitem [{\citenamefont {Stetcu}\ \emph {et~al.}(2020)\citenamefont {Stetcu},
  \citenamefont {Chadwick}, \citenamefont {Kawano}, \citenamefont {Talou},
  \citenamefont {Capote},\ and\ \citenamefont {Trkov}}]{Stetcu2020}%
  \BibitemOpen
  \bibfield  {author} {\bibinfo {author} {\bibfnamefont {I.}~\bibnamefont
  {Stetcu}}, \bibinfo {author} {\bibfnamefont {M.}~\bibnamefont {Chadwick}},
  \bibinfo {author} {\bibfnamefont {T.}~\bibnamefont {Kawano}}, \bibinfo
  {author} {\bibfnamefont {P.}~\bibnamefont {Talou}}, \bibinfo {author}
  {\bibfnamefont {R.}~\bibnamefont {Capote}}, \ and\ \bibinfo {author}
  {\bibfnamefont {A.}~\bibnamefont {Trkov}},\ }\href {\doibase
  https://doi.org/10.1016/j.nds.2019.12.007} {\bibfield  {journal} {\bibinfo
  {journal} {Nuclear Data Sheets}\ }\textbf {\bibinfo {volume} {163}},\
  \bibinfo {pages} {261} (\bibinfo {year} {2020})}\BibitemShut {NoStop}%
\bibitem [{\citenamefont {Bethe}(1936)}]{Bethe1936}%
  \BibitemOpen
  \bibfield  {author} {\bibinfo {author} {\bibfnamefont {H.~A.}\ \bibnamefont
  {Bethe}},\ }\href {\doibase 10.1103/PhysRev.50.332} {\bibfield  {journal}
  {\bibinfo  {journal} {Phys. Rev.}\ }\textbf {\bibinfo {volume} {50}},\
  \bibinfo {pages} {332} (\bibinfo {year} {1936})}\BibitemShut {NoStop}%
\bibitem [{\citenamefont {Kawano}(2019)}]{Kawano2019}%
  \BibitemOpen
  \bibfield  {author} {\bibinfo {author} {\bibfnamefont {T.}~\bibnamefont
  {Kawano}},\ }\href {\doibase 10.48550/ARXIV.1901.05641} {\enquote {\bibinfo
  {title} {Unified coupled-channels and hauser-feshbach model calculation for
  nuclear data evaluation},}\ } (\bibinfo {year} {2019})\BibitemShut {NoStop}%
\bibitem [{\citenamefont {Capote}\ \emph {et~al.}(2009)\citenamefont {Capote},
  \citenamefont {Herman}, \citenamefont {Obložinský}, \citenamefont {Young},
  \citenamefont {Goriely}, \citenamefont {Belgya}, \citenamefont {Ignatyuk},
  \citenamefont {Koning}, \citenamefont {Hilaire}, \citenamefont {Plujko},
  \citenamefont {Avrigeanu}, \citenamefont {Bersillon}, \citenamefont
  {Chadwick}, \citenamefont {Fukahori}, \citenamefont {Ge}, \citenamefont
  {Han}, \citenamefont {Kailas}, \citenamefont {Kopecky}, \citenamefont
  {Maslov}, \citenamefont {Reffo}, \citenamefont {Sin}, \citenamefont
  {Soukhovitskii},\ and\ \citenamefont {Talou}}]{Capote2009}%
  \BibitemOpen
  \bibfield  {author} {\bibinfo {author} {\bibfnamefont {R.}~\bibnamefont
  {Capote}}, \bibinfo {author} {\bibfnamefont {M.}~\bibnamefont {Herman}},
  \bibinfo {author} {\bibfnamefont {P.}~\bibnamefont {Obložinský}}, \bibinfo
  {author} {\bibfnamefont {P.}~\bibnamefont {Young}}, \bibinfo {author}
  {\bibfnamefont {S.}~\bibnamefont {Goriely}}, \bibinfo {author} {\bibfnamefont
  {T.}~\bibnamefont {Belgya}}, \bibinfo {author} {\bibfnamefont
  {A.}~\bibnamefont {Ignatyuk}}, \bibinfo {author} {\bibfnamefont
  {A.}~\bibnamefont {Koning}}, \bibinfo {author} {\bibfnamefont
  {S.}~\bibnamefont {Hilaire}}, \bibinfo {author} {\bibfnamefont
  {V.}~\bibnamefont {Plujko}}, \bibinfo {author} {\bibfnamefont
  {M.}~\bibnamefont {Avrigeanu}}, \bibinfo {author} {\bibfnamefont
  {O.}~\bibnamefont {Bersillon}}, \bibinfo {author} {\bibfnamefont
  {M.}~\bibnamefont {Chadwick}}, \bibinfo {author} {\bibfnamefont
  {T.}~\bibnamefont {Fukahori}}, \bibinfo {author} {\bibfnamefont
  {Z.}~\bibnamefont {Ge}}, \bibinfo {author} {\bibfnamefont {Y.}~\bibnamefont
  {Han}}, \bibinfo {author} {\bibfnamefont {S.}~\bibnamefont {Kailas}},
  \bibinfo {author} {\bibfnamefont {J.}~\bibnamefont {Kopecky}}, \bibinfo
  {author} {\bibfnamefont {V.}~\bibnamefont {Maslov}}, \bibinfo {author}
  {\bibfnamefont {G.}~\bibnamefont {Reffo}}, \bibinfo {author} {\bibfnamefont
  {M.}~\bibnamefont {Sin}}, \bibinfo {author} {\bibfnamefont {E.}~\bibnamefont
  {Soukhovitskii}}, \ and\ \bibinfo {author} {\bibfnamefont {P.}~\bibnamefont
  {Talou}},\ }\href {\doibase https://doi.org/10.1016/j.nds.2009.10.004}
  {\bibfield  {journal} {\bibinfo  {journal} {Nuclear Data Sheets}\ }\textbf
  {\bibinfo {volume} {110}},\ \bibinfo {pages} {3107} (\bibinfo {year}
  {2009})},\ \bibinfo {note} {special Issue on Nuclear Reaction
  Data}\BibitemShut {NoStop}%
\bibitem [{\citenamefont {{Thulliez, L.}}\ \emph {et~al.}(2016)\citenamefont
  {{Thulliez, L.}}, \citenamefont {{Litaize, O.}},\ and\ \citenamefont {{Serot,
  O.}}}]{Thulliez2016}%
  \BibitemOpen
  \bibfield  {author} {\bibinfo {author} {\bibnamefont {{Thulliez, L.}}},
  \bibinfo {author} {\bibnamefont {{Litaize, O.}}}, \ and\ \bibinfo {author}
  {\bibnamefont {{Serot, O.}}},\ }\href {\doibase 10.1051/epjconf/201611110003}
  {\bibfield  {journal} {\bibinfo  {journal} {EPJ Web of Conferences}\ }\textbf
  {\bibinfo {volume} {111}},\ \bibinfo {pages} {10003} (\bibinfo {year}
  {2016})}\BibitemShut {NoStop}%
\bibitem [{\citenamefont {Plompen}\ \emph {et~al.}(2020)\citenamefont
  {Plompen}, \citenamefont {Cabellos}, \citenamefont {De~Saint~Jean},
  \citenamefont {Fleming}, \citenamefont {Algora}, \citenamefont {Angelone},
  \citenamefont {Archier}, \citenamefont {Bauge}, \citenamefont {Bersillon},
  \citenamefont {Blokhin}, \citenamefont {Cantargi}, \citenamefont {Chebboubi},
  \citenamefont {Diez}, \citenamefont {Duarte}, \citenamefont {Dupont},
  \citenamefont {Dyrda}, \citenamefont {Erasmus}, \citenamefont {Fiorito},
  \citenamefont {Fischer}, \citenamefont {Flammini}, \citenamefont {Foligno},
  \citenamefont {Gilbert}, \citenamefont {Granada}, \citenamefont {Haeck},
  \citenamefont {Hambsch}, \citenamefont {Helgesson}, \citenamefont {Hilaire},
  \citenamefont {Hill}, \citenamefont {Hursin}, \citenamefont {Ichou},
  \citenamefont {Jacqmin}, \citenamefont {Jansky}, \citenamefont {Jouanne},
  \citenamefont {Kellett}, \citenamefont {Kim}, \citenamefont {Kim},
  \citenamefont {Kodeli}, \citenamefont {Koning}, \citenamefont {Konobeyev},
  \citenamefont {Kopecky}, \citenamefont {Kos}, \citenamefont {Kr{\'a}sa},
  \citenamefont {Leal}, \citenamefont {Leclaire}, \citenamefont {Leconte},
  \citenamefont {Lee}, \citenamefont {Leeb}, \citenamefont {Litaize},
  \citenamefont {Majerle}, \citenamefont {M{\'a}rquez Dami{\'a}n},
  \citenamefont {Michel-Sendis}, \citenamefont {Mills}, \citenamefont
  {Morillon}, \citenamefont {Nogu{\`e}re}, \citenamefont {Pecchia},
  \citenamefont {Pelloni}, \citenamefont {Pereslavtsev}, \citenamefont {Perry},
  \citenamefont {Rochman}, \citenamefont {R{\"o}hrmoser}, \citenamefont
  {Romain}, \citenamefont {Romojaro}, \citenamefont {Roubtsov}, \citenamefont
  {Sauvan}, \citenamefont {Schillebeeckx}, \citenamefont {Schmidt},
  \citenamefont {Serot}, \citenamefont {Simakov}, \citenamefont {Sirakov},
  \citenamefont {Sj{\"o}strand}, \citenamefont {Stankovskiy}, \citenamefont
  {Sublet}, \citenamefont {Tamagno}, \citenamefont {Trkov}, \citenamefont
  {van~der Marck}, \citenamefont {{\'A}lvarez-Velarde}, \citenamefont
  {Villari}, \citenamefont {Ware}, \citenamefont {Yokoyama},\ and\
  \citenamefont {{\v{Z}}erovnik}}]{Plompen2020}%
  \BibitemOpen
  \bibfield  {author} {\bibinfo {author} {\bibfnamefont {A.~J.~M.}\
  \bibnamefont {Plompen}}, \bibinfo {author} {\bibfnamefont {O.}~\bibnamefont
  {Cabellos}}, \bibinfo {author} {\bibfnamefont {C.}~\bibnamefont
  {De~Saint~Jean}}, \bibinfo {author} {\bibfnamefont {M.}~\bibnamefont
  {Fleming}}, \bibinfo {author} {\bibfnamefont {A.}~\bibnamefont {Algora}},
  \bibinfo {author} {\bibfnamefont {M.}~\bibnamefont {Angelone}}, \bibinfo
  {author} {\bibfnamefont {P.}~\bibnamefont {Archier}}, \bibinfo {author}
  {\bibfnamefont {E.}~\bibnamefont {Bauge}}, \bibinfo {author} {\bibfnamefont
  {O.}~\bibnamefont {Bersillon}}, \bibinfo {author} {\bibfnamefont
  {A.}~\bibnamefont {Blokhin}}, \bibinfo {author} {\bibfnamefont
  {F.}~\bibnamefont {Cantargi}}, \bibinfo {author} {\bibfnamefont
  {A.}~\bibnamefont {Chebboubi}}, \bibinfo {author} {\bibfnamefont
  {C.}~\bibnamefont {Diez}}, \bibinfo {author} {\bibfnamefont {H.}~\bibnamefont
  {Duarte}}, \bibinfo {author} {\bibfnamefont {E.}~\bibnamefont {Dupont}},
  \bibinfo {author} {\bibfnamefont {J.}~\bibnamefont {Dyrda}}, \bibinfo
  {author} {\bibfnamefont {B.}~\bibnamefont {Erasmus}}, \bibinfo {author}
  {\bibfnamefont {L.}~\bibnamefont {Fiorito}}, \bibinfo {author} {\bibfnamefont
  {U.}~\bibnamefont {Fischer}}, \bibinfo {author} {\bibfnamefont
  {D.}~\bibnamefont {Flammini}}, \bibinfo {author} {\bibfnamefont
  {D.}~\bibnamefont {Foligno}}, \bibinfo {author} {\bibfnamefont {M.~R.}\
  \bibnamefont {Gilbert}}, \bibinfo {author} {\bibfnamefont {J.~R.}\
  \bibnamefont {Granada}}, \bibinfo {author} {\bibfnamefont {W.}~\bibnamefont
  {Haeck}}, \bibinfo {author} {\bibfnamefont {F.-J.}\ \bibnamefont {Hambsch}},
  \bibinfo {author} {\bibfnamefont {P.}~\bibnamefont {Helgesson}}, \bibinfo
  {author} {\bibfnamefont {S.}~\bibnamefont {Hilaire}}, \bibinfo {author}
  {\bibfnamefont {I.}~\bibnamefont {Hill}}, \bibinfo {author} {\bibfnamefont
  {M.}~\bibnamefont {Hursin}}, \bibinfo {author} {\bibfnamefont
  {R.}~\bibnamefont {Ichou}}, \bibinfo {author} {\bibfnamefont
  {R.}~\bibnamefont {Jacqmin}}, \bibinfo {author} {\bibfnamefont
  {B.}~\bibnamefont {Jansky}}, \bibinfo {author} {\bibfnamefont
  {C.}~\bibnamefont {Jouanne}}, \bibinfo {author} {\bibfnamefont {M.~A.}\
  \bibnamefont {Kellett}}, \bibinfo {author} {\bibfnamefont {D.~H.}\
  \bibnamefont {Kim}}, \bibinfo {author} {\bibfnamefont {H.~I.}\ \bibnamefont
  {Kim}}, \bibinfo {author} {\bibfnamefont {I.}~\bibnamefont {Kodeli}},
  \bibinfo {author} {\bibfnamefont {A.~J.}\ \bibnamefont {Koning}}, \bibinfo
  {author} {\bibfnamefont {A.~Y.}\ \bibnamefont {Konobeyev}}, \bibinfo {author}
  {\bibfnamefont {S.}~\bibnamefont {Kopecky}}, \bibinfo {author} {\bibfnamefont
  {B.}~\bibnamefont {Kos}}, \bibinfo {author} {\bibfnamefont {A.}~\bibnamefont
  {Kr{\'a}sa}}, \bibinfo {author} {\bibfnamefont {L.~C.}\ \bibnamefont {Leal}},
  \bibinfo {author} {\bibfnamefont {N.}~\bibnamefont {Leclaire}}, \bibinfo
  {author} {\bibfnamefont {P.}~\bibnamefont {Leconte}}, \bibinfo {author}
  {\bibfnamefont {Y.~O.}\ \bibnamefont {Lee}}, \bibinfo {author} {\bibfnamefont
  {H.}~\bibnamefont {Leeb}}, \bibinfo {author} {\bibfnamefont {O.}~\bibnamefont
  {Litaize}}, \bibinfo {author} {\bibfnamefont {M.}~\bibnamefont {Majerle}},
  \bibinfo {author} {\bibfnamefont {J.~I.}\ \bibnamefont
  {M{\'a}rquez Dami{\'a}n}}, \bibinfo {author} {\bibfnamefont
  {F.}~\bibnamefont {Michel-Sendis}}, \bibinfo {author} {\bibfnamefont {R.~W.}\
  \bibnamefont {Mills}}, \bibinfo {author} {\bibfnamefont {B.}~\bibnamefont
  {Morillon}}, \bibinfo {author} {\bibfnamefont {G.}~\bibnamefont
  {Nogu{\`e}re}}, \bibinfo {author} {\bibfnamefont {M.}~\bibnamefont
  {Pecchia}}, \bibinfo {author} {\bibfnamefont {S.}~\bibnamefont {Pelloni}},
  \bibinfo {author} {\bibfnamefont {P.}~\bibnamefont {Pereslavtsev}}, \bibinfo
  {author} {\bibfnamefont {R.~J.}\ \bibnamefont {Perry}}, \bibinfo {author}
  {\bibfnamefont {D.}~\bibnamefont {Rochman}}, \bibinfo {author} {\bibfnamefont
  {A.}~\bibnamefont {R{\"o}hrmoser}}, \bibinfo {author} {\bibfnamefont
  {P.}~\bibnamefont {Romain}}, \bibinfo {author} {\bibfnamefont
  {P.}~\bibnamefont {Romojaro}}, \bibinfo {author} {\bibfnamefont
  {D.}~\bibnamefont {Roubtsov}}, \bibinfo {author} {\bibfnamefont
  {P.}~\bibnamefont {Sauvan}}, \bibinfo {author} {\bibfnamefont
  {P.}~\bibnamefont {Schillebeeckx}}, \bibinfo {author} {\bibfnamefont {K.~H.}\
  \bibnamefont {Schmidt}}, \bibinfo {author} {\bibfnamefont {O.}~\bibnamefont
  {Serot}}, \bibinfo {author} {\bibfnamefont {S.}~\bibnamefont {Simakov}},
  \bibinfo {author} {\bibfnamefont {I.}~\bibnamefont {Sirakov}}, \bibinfo
  {author} {\bibfnamefont {H.}~\bibnamefont {Sj{\"o}strand}}, \bibinfo {author}
  {\bibfnamefont {A.}~\bibnamefont {Stankovskiy}}, \bibinfo {author}
  {\bibfnamefont {J.~C.}\ \bibnamefont {Sublet}}, \bibinfo {author}
  {\bibfnamefont {P.}~\bibnamefont {Tamagno}}, \bibinfo {author} {\bibfnamefont
  {A.}~\bibnamefont {Trkov}}, \bibinfo {author} {\bibfnamefont
  {S.}~\bibnamefont {van~der Marck}}, \bibinfo {author} {\bibfnamefont
  {F.}~\bibnamefont {{\'A}lvarez-Velarde}}, \bibinfo {author} {\bibfnamefont
  {R.}~\bibnamefont {Villari}}, \bibinfo {author} {\bibfnamefont {T.~C.}\
  \bibnamefont {Ware}}, \bibinfo {author} {\bibfnamefont {K.}~\bibnamefont
  {Yokoyama}}, \ and\ \bibinfo {author} {\bibfnamefont {G.}~\bibnamefont
  {{\v{Z}}erovnik}},\ }\href {\doibase 10.1140/epja/s10050-020-00141-9}
  {\bibfield  {journal} {\bibinfo  {journal} {The European Physical Journal A}\
  }\textbf {\bibinfo {volume} {56}},\ \bibinfo {pages} {181} (\bibinfo {year}
  {2020})}\BibitemShut {NoStop}%
\bibitem [{\citenamefont {Bečvář}(1998)}]{Becvar1998}%
  \BibitemOpen
  \bibfield  {author} {\bibinfo {author} {\bibfnamefont {F.}~\bibnamefont
  {Bečvář}},\ }\href {\doibase
  https://doi.org/10.1016/S0168-9002(98)00787-6} {\bibfield  {journal}
  {\bibinfo  {journal} {Nuclear Instruments and Methods in Physics Research
  Section A: Accelerators, Spectrometers, Detectors and Associated Equipment}\
  }\textbf {\bibinfo {volume} {417}},\ \bibinfo {pages} {434} (\bibinfo {year}
  {1998})}\BibitemShut {NoStop}%
\bibitem [{\citenamefont {Regnier}\ \emph {et~al.}(2016)\citenamefont
  {Regnier}, \citenamefont {Litaize},\ and\ \citenamefont
  {Serot}}]{Regnier2016}%
  \BibitemOpen
  \bibfield  {author} {\bibinfo {author} {\bibfnamefont {D.}~\bibnamefont
  {Regnier}}, \bibinfo {author} {\bibfnamefont {O.}~\bibnamefont {Litaize}}, \
  and\ \bibinfo {author} {\bibfnamefont {O.}~\bibnamefont {Serot}},\ }\href
  {\doibase https://doi.org/10.1016/j.cpc.2015.12.007} {\bibfield  {journal}
  {\bibinfo  {journal} {Computer Physics Communications}\ }\textbf {\bibinfo
  {volume} {201}},\ \bibinfo {pages} {19} (\bibinfo {year} {2016})}\BibitemShut
  {NoStop}%
\bibitem [{\citenamefont {{Litaize, Olivier}}\ \emph
  {et~al.}(2017)\citenamefont {{Litaize, Olivier}}, \citenamefont {{Serot,
  Olivier}}, \citenamefont {{Thulliez, Lo\"{\i}c}},\ and\ \citenamefont
  {{Chebboubi, Abdelaziz}}}]{Litaize2017}%
  \BibitemOpen
  \bibfield  {author} {\bibinfo {author} {\bibnamefont {{Litaize, Olivier}}},
  \bibinfo {author} {\bibnamefont {{Serot, Olivier}}}, \bibinfo {author}
  {\bibnamefont {{Thulliez, Lo\"{\i}c}}}, \ and\ \bibinfo {author}
  {\bibnamefont {{Chebboubi, Abdelaziz}}},\ }\href {\doibase
  10.1051/epjconf/201714604006} {\bibfield  {journal} {\bibinfo  {journal} {EPJ
  Web Conf.}\ }\textbf {\bibinfo {volume} {146}},\ \bibinfo {pages} {04006}
  (\bibinfo {year} {2017})}\BibitemShut {NoStop}%
\bibitem [{\citenamefont {Litaize}\ \emph {et~al.}(2018)\citenamefont
  {Litaize}, \citenamefont {Chebboubi}, \citenamefont {Serot}, \citenamefont
  {Thulliez}, \citenamefont {Materna},\ and\ \citenamefont
  {Rapala}}]{Litaize2018InfluenceStructure}%
  \BibitemOpen
  \bibfield  {author} {\bibinfo {author} {\bibfnamefont {O.}~\bibnamefont
  {Litaize}}, \bibinfo {author} {\bibfnamefont {A.}~\bibnamefont {Chebboubi}},
  \bibinfo {author} {\bibfnamefont {O.}~\bibnamefont {Serot}}, \bibinfo
  {author} {\bibfnamefont {L.}~\bibnamefont {Thulliez}}, \bibinfo {author}
  {\bibfnamefont {T.}~\bibnamefont {Materna}}, \ and\ \bibinfo {author}
  {\bibfnamefont {M.}~\bibnamefont {Rapala}},\ }\href {\doibase
  10.1051/epjn/2018043} {\bibfield  {journal} {\bibinfo  {journal} {EPJ Nuclear
  Sci. Technol.}\ }\textbf {\bibinfo {volume} {4}},\ \bibinfo {pages} {28}
  (\bibinfo {year} {2018})}\BibitemShut {NoStop}%
\bibitem [{\citenamefont {Randrup}\ and\ \citenamefont
  {Vogt}(2014)}]{Randrup2014}%
  \BibitemOpen
  \bibfield  {author} {\bibinfo {author} {\bibfnamefont {J.}~\bibnamefont
  {Randrup}}\ and\ \bibinfo {author} {\bibfnamefont {R.}~\bibnamefont {Vogt}},\
  }\href {\doibase 10.1103/PhysRevC.89.044601} {\bibfield  {journal} {\bibinfo
  {journal} {Phys. Rev. C}\ }\textbf {\bibinfo {volume} {89}},\ \bibinfo
  {pages} {044601} (\bibinfo {year} {2014})}\BibitemShut {NoStop}%
\bibitem [{\citenamefont {Vogt}\ and\ \citenamefont
  {Randrup}(2017)}]{Vogt2017}%
  \BibitemOpen
  \bibfield  {author} {\bibinfo {author} {\bibfnamefont {R.}~\bibnamefont
  {Vogt}}\ and\ \bibinfo {author} {\bibfnamefont {J.}~\bibnamefont {Randrup}},\
  }\href {\doibase 10.1103/PhysRevC.96.064620} {\bibfield  {journal} {\bibinfo
  {journal} {Phys. Rev. C}\ }\textbf {\bibinfo {volume} {96}},\ \bibinfo
  {pages} {064620} (\bibinfo {year} {2017})}\BibitemShut {NoStop}%
\bibitem [{\citenamefont {Gatera}\ \emph {et~al.}(2017)\citenamefont {Gatera},
  \citenamefont {Belgya}, \citenamefont {Geerts}, \citenamefont {G\"o\"ok},
  \citenamefont {Hambsch}, \citenamefont {Lebois}, \citenamefont {Mar\'oti},
  \citenamefont {Moens}, \citenamefont {Oberstedt}, \citenamefont {Oberstedt},
  \citenamefont {Postelt}, \citenamefont {Qi}, \citenamefont {Szentmikl\'osi},
  \citenamefont {Sibbens}, \citenamefont {Vanleeuw}, \citenamefont {Vidali},\
  and\ \citenamefont {Zeiser}}]{Gatera2017}%
  \BibitemOpen
  \bibfield  {author} {\bibinfo {author} {\bibfnamefont {A.}~\bibnamefont
  {Gatera}}, \bibinfo {author} {\bibfnamefont {T.}~\bibnamefont {Belgya}},
  \bibinfo {author} {\bibfnamefont {W.}~\bibnamefont {Geerts}}, \bibinfo
  {author} {\bibfnamefont {A.}~\bibnamefont {G\"o\"ok}}, \bibinfo {author}
  {\bibfnamefont {F.-J.}\ \bibnamefont {Hambsch}}, \bibinfo {author}
  {\bibfnamefont {M.}~\bibnamefont {Lebois}}, \bibinfo {author} {\bibfnamefont
  {B.}~\bibnamefont {Mar\'oti}}, \bibinfo {author} {\bibfnamefont
  {A.}~\bibnamefont {Moens}}, \bibinfo {author} {\bibfnamefont
  {A.}~\bibnamefont {Oberstedt}}, \bibinfo {author} {\bibfnamefont
  {S.}~\bibnamefont {Oberstedt}}, \bibinfo {author} {\bibfnamefont
  {F.}~\bibnamefont {Postelt}}, \bibinfo {author} {\bibfnamefont
  {L.}~\bibnamefont {Qi}}, \bibinfo {author} {\bibfnamefont {L.}~\bibnamefont
  {Szentmikl\'osi}}, \bibinfo {author} {\bibfnamefont {G.}~\bibnamefont
  {Sibbens}}, \bibinfo {author} {\bibfnamefont {D.}~\bibnamefont {Vanleeuw}},
  \bibinfo {author} {\bibfnamefont {M.}~\bibnamefont {Vidali}}, \ and\ \bibinfo
  {author} {\bibfnamefont {F.}~\bibnamefont {Zeiser}},\ }\href {\doibase
  https://doi.org/10.1103/PhysRevC.95.064609} {\bibfield  {journal} {\bibinfo
  {journal} {Phys. Rev. C}\ }\textbf {\bibinfo {volume} {95}},\ \bibinfo
  {pages} {064609} (\bibinfo {year} {2017})}\BibitemShut {NoStop}%
\bibitem [{\citenamefont {Chadwick}\ \emph {et~al.}(2011)\citenamefont
  {Chadwick}, \citenamefont {Herman}, \citenamefont {Obložinský},
  \citenamefont {Dunn}, \citenamefont {Danon}, \citenamefont {Kahler},
  \citenamefont {Smith}, \citenamefont {Pritychenko}, \citenamefont {Arbanas},
  \citenamefont {Arcilla}, \citenamefont {Brewer}, \citenamefont {Brown},
  \citenamefont {Capote}, \citenamefont {Carlson}, \citenamefont {Cho},
  \citenamefont {Derrien}, \citenamefont {Guber}, \citenamefont {Hale},
  \citenamefont {Hoblit}, \citenamefont {Holloway}, \citenamefont {Johnson},
  \citenamefont {Kawano}, \citenamefont {Kiedrowski}, \citenamefont {Kim},
  \citenamefont {Kunieda}, \citenamefont {Larson}, \citenamefont {Leal},
  \citenamefont {Lestone}, \citenamefont {Little}, \citenamefont {McCutchan},
  \citenamefont {MacFarlane}, \citenamefont {MacInnes}, \citenamefont
  {Mattoon}, \citenamefont {McKnight}, \citenamefont {Mughabghab},
  \citenamefont {Nobre}, \citenamefont {Palmiotti}, \citenamefont {Palumbo},
  \citenamefont {Pigni}, \citenamefont {Pronyaev}, \citenamefont {Sayer},
  \citenamefont {Sonzogni}, \citenamefont {Summers}, \citenamefont {Talou},
  \citenamefont {Thompson}, \citenamefont {Trkov}, \citenamefont {Vogt},
  \citenamefont {{van der Marck}}, \citenamefont {Wallner}, \citenamefont
  {White}, \citenamefont {Wiarda},\ and\ \citenamefont {Young}}]{Chadwick2011}%
  \BibitemOpen
  \bibfield  {author} {\bibinfo {author} {\bibfnamefont {M.}~\bibnamefont
  {Chadwick}}, \bibinfo {author} {\bibfnamefont {M.}~\bibnamefont {Herman}},
  \bibinfo {author} {\bibfnamefont {P.}~\bibnamefont {Obložinský}}, \bibinfo
  {author} {\bibfnamefont {M.}~\bibnamefont {Dunn}}, \bibinfo {author}
  {\bibfnamefont {Y.}~\bibnamefont {Danon}}, \bibinfo {author} {\bibfnamefont
  {A.}~\bibnamefont {Kahler}}, \bibinfo {author} {\bibfnamefont
  {D.}~\bibnamefont {Smith}}, \bibinfo {author} {\bibfnamefont
  {B.}~\bibnamefont {Pritychenko}}, \bibinfo {author} {\bibfnamefont
  {G.}~\bibnamefont {Arbanas}}, \bibinfo {author} {\bibfnamefont
  {R.}~\bibnamefont {Arcilla}}, \bibinfo {author} {\bibfnamefont
  {R.}~\bibnamefont {Brewer}}, \bibinfo {author} {\bibfnamefont
  {D.}~\bibnamefont {Brown}}, \bibinfo {author} {\bibfnamefont
  {R.}~\bibnamefont {Capote}}, \bibinfo {author} {\bibfnamefont
  {A.}~\bibnamefont {Carlson}}, \bibinfo {author} {\bibfnamefont
  {Y.}~\bibnamefont {Cho}}, \bibinfo {author} {\bibfnamefont {H.}~\bibnamefont
  {Derrien}}, \bibinfo {author} {\bibfnamefont {K.}~\bibnamefont {Guber}},
  \bibinfo {author} {\bibfnamefont {G.}~\bibnamefont {Hale}}, \bibinfo {author}
  {\bibfnamefont {S.}~\bibnamefont {Hoblit}}, \bibinfo {author} {\bibfnamefont
  {S.}~\bibnamefont {Holloway}}, \bibinfo {author} {\bibfnamefont
  {T.}~\bibnamefont {Johnson}}, \bibinfo {author} {\bibfnamefont
  {T.}~\bibnamefont {Kawano}}, \bibinfo {author} {\bibfnamefont
  {B.}~\bibnamefont {Kiedrowski}}, \bibinfo {author} {\bibfnamefont
  {H.}~\bibnamefont {Kim}}, \bibinfo {author} {\bibfnamefont {S.}~\bibnamefont
  {Kunieda}}, \bibinfo {author} {\bibfnamefont {N.}~\bibnamefont {Larson}},
  \bibinfo {author} {\bibfnamefont {L.}~\bibnamefont {Leal}}, \bibinfo {author}
  {\bibfnamefont {J.}~\bibnamefont {Lestone}}, \bibinfo {author} {\bibfnamefont
  {R.}~\bibnamefont {Little}}, \bibinfo {author} {\bibfnamefont
  {E.}~\bibnamefont {McCutchan}}, \bibinfo {author} {\bibfnamefont
  {R.}~\bibnamefont {MacFarlane}}, \bibinfo {author} {\bibfnamefont
  {M.}~\bibnamefont {MacInnes}}, \bibinfo {author} {\bibfnamefont
  {C.}~\bibnamefont {Mattoon}}, \bibinfo {author} {\bibfnamefont
  {R.}~\bibnamefont {McKnight}}, \bibinfo {author} {\bibfnamefont
  {S.}~\bibnamefont {Mughabghab}}, \bibinfo {author} {\bibfnamefont
  {G.}~\bibnamefont {Nobre}}, \bibinfo {author} {\bibfnamefont
  {G.}~\bibnamefont {Palmiotti}}, \bibinfo {author} {\bibfnamefont
  {A.}~\bibnamefont {Palumbo}}, \bibinfo {author} {\bibfnamefont
  {M.}~\bibnamefont {Pigni}}, \bibinfo {author} {\bibfnamefont
  {V.}~\bibnamefont {Pronyaev}}, \bibinfo {author} {\bibfnamefont
  {R.}~\bibnamefont {Sayer}}, \bibinfo {author} {\bibfnamefont
  {A.}~\bibnamefont {Sonzogni}}, \bibinfo {author} {\bibfnamefont
  {N.}~\bibnamefont {Summers}}, \bibinfo {author} {\bibfnamefont
  {P.}~\bibnamefont {Talou}}, \bibinfo {author} {\bibfnamefont
  {I.}~\bibnamefont {Thompson}}, \bibinfo {author} {\bibfnamefont
  {A.}~\bibnamefont {Trkov}}, \bibinfo {author} {\bibfnamefont
  {R.}~\bibnamefont {Vogt}}, \bibinfo {author} {\bibfnamefont {S.}~\bibnamefont
  {{van der Marck}}}, \bibinfo {author} {\bibfnamefont {A.}~\bibnamefont
  {Wallner}}, \bibinfo {author} {\bibfnamefont {M.}~\bibnamefont {White}},
  \bibinfo {author} {\bibfnamefont {D.}~\bibnamefont {Wiarda}}, \ and\ \bibinfo
  {author} {\bibfnamefont {P.}~\bibnamefont {Young}},\ }\href {\doibase
  https://doi.org/10.1016/j.nds.2011.11.002} {\bibfield  {journal} {\bibinfo
  {journal} {Nuclear Data Sheets}\ }\textbf {\bibinfo {volume} {112}},\
  \bibinfo {pages} {2887} (\bibinfo {year} {2011})},\ \bibinfo {note} {special
  Issue on ENDF/B-VII.1 Library}\BibitemShut {NoStop}%
\bibitem [{\citenamefont {Mumpower}\ \emph {et~al.}(2016)\citenamefont
  {Mumpower}, \citenamefont {Kawano},\ and\ \citenamefont
  {M\"oller}}]{Mumpower2016}%
  \BibitemOpen
  \bibfield  {author} {\bibinfo {author} {\bibfnamefont {M.~R.}\ \bibnamefont
  {Mumpower}}, \bibinfo {author} {\bibfnamefont {T.}~\bibnamefont {Kawano}}, \
  and\ \bibinfo {author} {\bibfnamefont {P.}~\bibnamefont {M\"oller}},\ }\href
  {\doibase 10.1103/PhysRevC.94.064317} {\bibfield  {journal} {\bibinfo
  {journal} {Phys. Rev. C}\ }\textbf {\bibinfo {volume} {94}},\ \bibinfo
  {pages} {064317} (\bibinfo {year} {2016})}\BibitemShut {NoStop}%
\bibitem [{\citenamefont {Spyrou}\ \emph {et~al.}(2016)\citenamefont {Spyrou},
  \citenamefont {Liddick}, \citenamefont {Naqvi}, \citenamefont {Crider},
  \citenamefont {Dombos}, \citenamefont {Bleuel}, \citenamefont {Brown},
  \citenamefont {Couture}, \citenamefont {Crespo~Campo}, \citenamefont
  {Guttormsen}, \citenamefont {Larsen}, \citenamefont {Lewis}, \citenamefont
  {M\"oller}, \citenamefont {Mosby}, \citenamefont {Mumpower}, \citenamefont
  {Perdikakis}, \citenamefont {Prokop}, \citenamefont {Renstr\o{}m},
  \citenamefont {Siem}, \citenamefont {Quinn},\ and\ \citenamefont
  {Valenta}}]{Spyrou2016}%
  \BibitemOpen
  \bibfield  {author} {\bibinfo {author} {\bibfnamefont {A.}~\bibnamefont
  {Spyrou}}, \bibinfo {author} {\bibfnamefont {S.~N.}\ \bibnamefont {Liddick}},
  \bibinfo {author} {\bibfnamefont {F.}~\bibnamefont {Naqvi}}, \bibinfo
  {author} {\bibfnamefont {B.~P.}\ \bibnamefont {Crider}}, \bibinfo {author}
  {\bibfnamefont {A.~C.}\ \bibnamefont {Dombos}}, \bibinfo {author}
  {\bibfnamefont {D.~L.}\ \bibnamefont {Bleuel}}, \bibinfo {author}
  {\bibfnamefont {B.~A.}\ \bibnamefont {Brown}}, \bibinfo {author}
  {\bibfnamefont {A.}~\bibnamefont {Couture}}, \bibinfo {author} {\bibfnamefont
  {L.}~\bibnamefont {Crespo~Campo}}, \bibinfo {author} {\bibfnamefont
  {M.}~\bibnamefont {Guttormsen}}, \bibinfo {author} {\bibfnamefont {A.~C.}\
  \bibnamefont {Larsen}}, \bibinfo {author} {\bibfnamefont {R.}~\bibnamefont
  {Lewis}}, \bibinfo {author} {\bibfnamefont {P.}~\bibnamefont {M\"oller}},
  \bibinfo {author} {\bibfnamefont {S.}~\bibnamefont {Mosby}}, \bibinfo
  {author} {\bibfnamefont {M.~R.}\ \bibnamefont {Mumpower}}, \bibinfo {author}
  {\bibfnamefont {G.}~\bibnamefont {Perdikakis}}, \bibinfo {author}
  {\bibfnamefont {C.~J.}\ \bibnamefont {Prokop}}, \bibinfo {author}
  {\bibfnamefont {T.}~\bibnamefont {Renstr\o{}m}}, \bibinfo {author}
  {\bibfnamefont {S.}~\bibnamefont {Siem}}, \bibinfo {author} {\bibfnamefont
  {S.~J.}\ \bibnamefont {Quinn}}, \ and\ \bibinfo {author} {\bibfnamefont
  {S.}~\bibnamefont {Valenta}},\ }\href {\doibase
  https://doi.org/10.1103/PhysRevLett.117.142701} {\bibfield  {journal}
  {\bibinfo  {journal} {Phys. Rev. Lett.}\ }\textbf {\bibinfo {volume} {117}},\
  \bibinfo {pages} {142701} (\bibinfo {year} {2016})}\BibitemShut {NoStop}%
\bibitem [{\citenamefont {Boutoux}\ \emph {et~al.}(2012)\citenamefont
  {Boutoux}, \citenamefont {Jurado}, \citenamefont {Méot}, \citenamefont
  {Roig}, \citenamefont {Mathieu}, \citenamefont {Aïche}, \citenamefont
  {Barreau}, \citenamefont {Capellan}, \citenamefont {Companis}, \citenamefont
  {Czajkowski}, \citenamefont {Schmidt}, \citenamefont {Burke}, \citenamefont
  {Bail}, \citenamefont {Daugas}, \citenamefont {Faul}, \citenamefont {Morel},
  \citenamefont {Pillet}, \citenamefont {Théroine}, \citenamefont {Derkx},
  \citenamefont {Sérot}, \citenamefont {Matéa},\ and\ \citenamefont
  {Tassan-Got}}]{Boutoux2012}%
  \BibitemOpen
  \bibfield  {author} {\bibinfo {author} {\bibfnamefont {G.}~\bibnamefont
  {Boutoux}}, \bibinfo {author} {\bibfnamefont {B.}~\bibnamefont {Jurado}},
  \bibinfo {author} {\bibfnamefont {V.}~\bibnamefont {Méot}}, \bibinfo
  {author} {\bibfnamefont {O.}~\bibnamefont {Roig}}, \bibinfo {author}
  {\bibfnamefont {L.}~\bibnamefont {Mathieu}}, \bibinfo {author} {\bibfnamefont
  {M.}~\bibnamefont {Aïche}}, \bibinfo {author} {\bibfnamefont
  {G.}~\bibnamefont {Barreau}}, \bibinfo {author} {\bibfnamefont
  {N.}~\bibnamefont {Capellan}}, \bibinfo {author} {\bibfnamefont
  {I.}~\bibnamefont {Companis}}, \bibinfo {author} {\bibfnamefont
  {S.}~\bibnamefont {Czajkowski}}, \bibinfo {author} {\bibfnamefont {K.-H.}\
  \bibnamefont {Schmidt}}, \bibinfo {author} {\bibfnamefont {J.}~\bibnamefont
  {Burke}}, \bibinfo {author} {\bibfnamefont {A.}~\bibnamefont {Bail}},
  \bibinfo {author} {\bibfnamefont {J.}~\bibnamefont {Daugas}}, \bibinfo
  {author} {\bibfnamefont {T.}~\bibnamefont {Faul}}, \bibinfo {author}
  {\bibfnamefont {P.}~\bibnamefont {Morel}}, \bibinfo {author} {\bibfnamefont
  {N.}~\bibnamefont {Pillet}}, \bibinfo {author} {\bibfnamefont
  {C.}~\bibnamefont {Théroine}}, \bibinfo {author} {\bibfnamefont
  {X.}~\bibnamefont {Derkx}}, \bibinfo {author} {\bibfnamefont
  {O.}~\bibnamefont {Sérot}}, \bibinfo {author} {\bibfnamefont
  {I.}~\bibnamefont {Matéa}}, \ and\ \bibinfo {author} {\bibfnamefont
  {L.}~\bibnamefont {Tassan-Got}},\ }\href {\doibase
  https://doi.org/10.1016/j.physletb.2012.05.012} {\bibfield  {journal}
  {\bibinfo  {journal} {Physics Letters B}\ }\textbf {\bibinfo {volume}
  {712}},\ \bibinfo {pages} {319} (\bibinfo {year} {2012})}\BibitemShut
  {NoStop}%
\bibitem [{\citenamefont {Zeiser}\ \emph {et~al.}(2019)\citenamefont {Zeiser},
  \citenamefont {Tveten}, \citenamefont {Potel}, \citenamefont {Larsen},
  \citenamefont {Guttormsen}, \citenamefont {Laplace}, \citenamefont {Siem},
  \citenamefont {Bleuel}, \citenamefont {Goldblum}, \citenamefont {Bernstein},
  \citenamefont {Bello~Garrote}, \citenamefont {Crespo~Campo}, \citenamefont
  {Eriksen}, \citenamefont {G\"orgen}, \citenamefont {Hadynska-Klek},
  \citenamefont {Ingeberg}, \citenamefont {Midtb\o{}}, \citenamefont {Sahin},
  \citenamefont {Tornyi}, \citenamefont {Voinov}, \citenamefont {Wiedeking},\
  and\ \citenamefont {Wilson}}]{Zeiser2019}%
  \BibitemOpen
  \bibfield  {author} {\bibinfo {author} {\bibfnamefont {F.}~\bibnamefont
  {Zeiser}}, \bibinfo {author} {\bibfnamefont {G.~M.}\ \bibnamefont {Tveten}},
  \bibinfo {author} {\bibfnamefont {G.}~\bibnamefont {Potel}}, \bibinfo
  {author} {\bibfnamefont {A.~C.}\ \bibnamefont {Larsen}}, \bibinfo {author}
  {\bibfnamefont {M.}~\bibnamefont {Guttormsen}}, \bibinfo {author}
  {\bibfnamefont {T.~A.}\ \bibnamefont {Laplace}}, \bibinfo {author}
  {\bibfnamefont {S.}~\bibnamefont {Siem}}, \bibinfo {author} {\bibfnamefont
  {D.~L.}\ \bibnamefont {Bleuel}}, \bibinfo {author} {\bibfnamefont {B.~L.}\
  \bibnamefont {Goldblum}}, \bibinfo {author} {\bibfnamefont {L.~A.}\
  \bibnamefont {Bernstein}}, \bibinfo {author} {\bibfnamefont {F.~L.}\
  \bibnamefont {Bello~Garrote}}, \bibinfo {author} {\bibfnamefont
  {L.}~\bibnamefont {Crespo~Campo}}, \bibinfo {author} {\bibfnamefont {T.~K.}\
  \bibnamefont {Eriksen}}, \bibinfo {author} {\bibfnamefont {A.}~\bibnamefont
  {G\"orgen}}, \bibinfo {author} {\bibfnamefont {K.}~\bibnamefont
  {Hadynska-Klek}}, \bibinfo {author} {\bibfnamefont {V.~W.}\ \bibnamefont
  {Ingeberg}}, \bibinfo {author} {\bibfnamefont {J.~E.}\ \bibnamefont
  {Midtb\o{}}}, \bibinfo {author} {\bibfnamefont {E.}~\bibnamefont {Sahin}},
  \bibinfo {author} {\bibfnamefont {T.}~\bibnamefont {Tornyi}}, \bibinfo
  {author} {\bibfnamefont {A.}~\bibnamefont {Voinov}}, \bibinfo {author}
  {\bibfnamefont {M.}~\bibnamefont {Wiedeking}}, \ and\ \bibinfo {author}
  {\bibfnamefont {J.}~\bibnamefont {Wilson}},\ }\href {\doibase
  https://doi.org/10.1103/PhysRevC.100.024305} {\bibfield  {journal} {\bibinfo
  {journal} {Phys. Rev. C}\ }\textbf {\bibinfo {volume} {100}},\ \bibinfo
  {pages} {024305} (\bibinfo {year} {2019})}\BibitemShut {NoStop}%
\bibitem [{\citenamefont {Marin}\ \emph {et~al.}(2022)\citenamefont {Marin},
  \citenamefont {Sansevero}, \citenamefont {Okar}, \citenamefont {Hernandez},
  \citenamefont {Vogt}, \citenamefont {Randrup}, \citenamefont {Clarke},
  \citenamefont {Protopopescu},\ and\ \citenamefont {Pozzi}}]{Marin2022}%
  \BibitemOpen
  \bibfield  {author} {\bibinfo {author} {\bibfnamefont {S.}~\bibnamefont
  {Marin}}, \bibinfo {author} {\bibfnamefont {E.~P.}\ \bibnamefont
  {Sansevero}}, \bibinfo {author} {\bibfnamefont {M.~S.}\ \bibnamefont {Okar}},
  \bibinfo {author} {\bibfnamefont {I.~E.}\ \bibnamefont {Hernandez}}, \bibinfo
  {author} {\bibfnamefont {R.}~\bibnamefont {Vogt}}, \bibinfo {author}
  {\bibfnamefont {J.}~\bibnamefont {Randrup}}, \bibinfo {author} {\bibfnamefont
  {S.~D.}\ \bibnamefont {Clarke}}, \bibinfo {author} {\bibfnamefont {V.~A.}\
  \bibnamefont {Protopopescu}}, \ and\ \bibinfo {author} {\bibfnamefont
  {S.~A.}\ \bibnamefont {Pozzi}},\ }\href {\doibase
  https://doi.org/10.1103/PhysRevC.105.054609} {\bibfield  {journal} {\bibinfo
  {journal} {Phys. Rev. C}\ }\textbf {\bibinfo {volume} {105}},\ \bibinfo
  {pages} {054609} (\bibinfo {year} {2022})}\BibitemShut {NoStop}%
\bibitem [{NuD()}]{NuDat}%
  \BibitemOpen
  \href@noop {} {}\bibinfo {howpublished}
  {\url{https://www.nndc.bnl.gov/nudat3/}}\BibitemShut {NoStop}%
\end{thebibliography}%


%


\end{document}